\def\galex{{\sl GALEX}}
\def\gemini{{\sl Gemini}}
\def\hst{{\sl HST}}

\def\acs{{\sl ACS}}
\def\wfc3{{\sl WFC3}}
\def\spitzer{{\sl Spitzer}}

\def\herschel {{\sl Herschel}}

\documentclass[12pt,preprint]{aastex}
\usepackage{epsf}
\usepackage{color}
\usepackage{graphicx}
\usepackage{graphics}
\usepackage{epsfig}
\usepackage{longtable}
\usepackage{multirow}
\begin{document}
\title{The Star Formation History in the M31 Bulge}
\author{Hui Dong$^{1}$, Knut Olsen$^2$, Tod Lauer$^2$, Abhijit Saha$^2$, 
Zhiyuan Li$^{3,4}$, Ruben Garc\'ia-Benito$^{1}$, Rainer Sch{\"o}del$^1$}

\affil{$^1$ Instituto de Astrof\'{i}sica de Andaluc\'{i}a (CSIC), 
Glorieta de la Astronom\'{a} S/N, E-18008 Granada, Spain}
\affil{$^2$ National Optical Astronomy Observatory,
Tucson, AZ, 85719, USA}\affil{$^3$ School of Astronomy and Space 
Science, Nanjing University, Nanjing, 210093, China}\affil{$^4$ Key Laboratory of Modern Astronomy and Astrophysics at Nanjing University, Ministry of Education, Nanjing 210093, China}\affil{E-mail: hdong@iaa.es}

\begin{abstract}
We present the study of stellar populations in the 
central 5.5\arcmin\ ($\sim$1.2 kpc) of the M31 bulge by using the optical 
color magnitude 
diagram derived from HST ACS WFC/HRC observations.  
%from F330W (3354\AA ) to 
%F814W (8333\AA ) bands. 
In order to 
enhance image quality and then obtain deeper photometry, 
we construct Nyquist-sampled images and use a deconvolution 
method to detect sources and measure their photometry. 
We demonstrate that our method performs better than 
{\tt DOLPHOT} in the extremely crowded region. 
The resolved stars in 
the M31 bulge have been divided into nine annuli and the color magnitude 
diagram fitting is performed 
for each of them. % to study the evolution of the properties of the stellar populations 
%along the galactocentric radius in the M31 bulge. 
We confirm that the 
majority of stars ($>$70\%) in the M31 bulge 
are indeed very old ($>$ 5 Gyr) 
and metal-rich ([Fe/H]$\sim$0.3). At later times, the star 
formation rate decreased and then experienced  
a significant rise around 1 Gyr ago, which pervaded the entire 
M31 bulge. After that, stars formed at less than 500 Myr ago in 
the central 130\arcsec . Through simulation, we find that these intermediate-age stars 
cannot be the artifacts introduced by the blending effect. Our results suggest that 
although the majority of the M31 bulge %formed though major mergers, 
are very old, the 
secular evolutionary process still continuously builds up the M31 bulge slowly. We compare our 
star formation history with an older analysis derived from the spectral energy distribution 
fitting, which suggests that the latter one is still a reasonable tool 
for the study of stellar populations in remote galaxies. 


%  although not too much, 
%which contributes less than 3\% 
%of stellar mass, which strongly against the scenario of the M31 and M32 collision. 

%This result suggest that the M31 bulge experienced several 
%star formation activities recently.  
\end{abstract}

\section{Introduction}
%Bulges are a unique component in spiral galaxies. Their surface brightness is 
The surface brightness of bulges in spiral galaxies is 
smooth and their morphologies are similar to elliptical galaxies. 
Therefore, at first, bulges were thought to form through the same mechanism as  
elliptical galaxies, 
such as rapid collapse in the early Universe~\citep{san90} or major 
mergers~\citep{too77},  
which remove molecular clouds, stop star formation and increase 
velocity dispersion. In both cases, old stars (at least several Gyr old) are 
expected. In the former case, the stellar metallicity should be very low, while the 
metallicity in the latter cases depends on when the major mergers happened. 
On the other hand, 
young stars have also been found in bulges and are 
explained by another mechanism, secular evolutionary process, which 
could also play a significant role in shaping bulges; Materials in the disks spiral 
into inner regions of galaxies, assisted by the torque of galactic bars, trigger star formation and 
produce young stars~\citep[see the excellent review given by][]{kor04}. Key to disentangle these  mechanisms is to study the spatial 
distribution of the age and metallicity of stars in spiral bulges. 

The Andromeda galaxy (M31) is the closest external grand design spiral 
galaxy (780 kpc,~\citealt{mcc05,gri14})     
and has a distinct bulge~\citep{bea07}, as well as a super-massive black 
hole~\citep[M31*,][]{dre88,kor88,cra92,gar10,li11}. 
The S\'ersic index of the M31 bulge at the $I$ band is 2.2~\citep{cou11}. Therefore, 
the M31 bulge is classified as a classical bulge according to the criterium given 
in~\citet{fis10}. On the other hand,~\citet{bea07} identified a thin bar in the M31 bulge 
(see also~\citealt{bla17,opi18}), 
which is the major characterization of pseudobulges~\citep{kor04}. 
Compared to our own Milky Way, M31 is a better lab to 
study the formation and evolution history of bulges for two reasons: First, the M31 bulge 
suffers from small foreground extinction from the Milky Way 
($A_V$$\sim$0.166 mag,~\citealt{sch11}), because of its Galactic latitude (-21.57 degree), and 
local extinction from a small amount of molecular clouds in the M31 
bulge~\citep{mel13,don16}. Therefore, unlike the Galactic bulge, the systematic uncertainty 
introduced by extinction can be ignored. Second, the M31 bulge is 
distant enough that we do not need to consider the differential distances   
of individual stars, while the M31 bulge is still close enough that we can resolve 
individual bright stars. 

The biggest challenge to characterize the stellar populations 
in the M31 bulge is the crowding due to high surface 
brightness~\citep{ric93,dep93,ric95,dav01,ste03}.
\citet{ste03} analyzed the observation taken by the 
Hubble Space Telescope (\hst ) 
of the M31 bulge with an angular resolution 
of 0.2\arcsec (0.8 pc). They found that their stars 
were significantly dimmer ($\sim$ 1 mag) than the ones reported 
in~\citet{ric95}, which used the \hst\ dataset 
with the original aberrated optics, and in~\citet{dav01}, which used the 3.6 m 
Canada-France-Hawaii Telescope with a seeing of 
0.35\arcsec\ (1.3 pc). This magnitude difference was apparently caused by the 
blending of multiple sources in the observation with poor angular resolutions. 

%They found that the brightest 
%asymptotic giant branch stars (AGB) is 1.5 mag dimmer at the $K$ band 
%than that found by~\citet{ric95}, which use the \hst\ dataset 
%with the original aberrated optics, and %, 
%such as Red Giant Branch (RGB) stars (see \S\ref{???}). 
%For example,~\citet{ste03} suggested that 
%the blending could cause as high as 0.75 mag and 0.55 mag brightening at 
%the $J$ and $K$ bands.

In order to overcome the crowding problem,~\citet{ols06} (see also~\citealt{dav05}) 
analyzed adaptive optic (AO) images of the M31 bulge taken by the \gemini/NIRI camera 
in the near-IR (NIR) band, with an angular resolution of 
$\sim$0.1\arcsec\ (0.4 pc). Although their 
observation was still not deep enough to reach red clump (RC) 
stars, they detected a significant amount of red giant branch (RGB) stars 
and asymptotic giant branch (AGB) stars. Through fitting of the 
NIR color magnitude diagram (CMD), they found that stars with 
$\geq$ 6 Gyr and solar metallicity contribute significantly to the M31 bulge 
($\geq$60\%).~\citet{sag10} analyzed optical low resolution spectra taken 
by {\sl Hobby-Eberly} Telescope and derived similar results, while they suggested  
an even older age (12 Gyr). % for the old stellar population. 
They also found that 
the stellar metallicity decreased from 3$Z_{\odot}$ in the nucleus to 
1$Z_{\odot}$ at 2\arcmin\ (454 pc) away from the centre.~\citet{don15} performed  
the spectral energy distribution (SED) fitting of the photometric data derived 
from \hst\ images at ten bands from ultraviolet (UV) to NIR and obtained similar 
results. Meanwhile, no stellar population younger than 10 Myr seems to 
exist in the M31 bulge~\citep{ros12}. 
 
%The M31 bulge is known to be dominated by old and metal-rich 
%stars. Although earlier stud and HST/NICMOS data and claim that 

%, plus a very small 
%fraction of intermediate-age stars.

On the other hand, ~\citet{ben05} reported that the \hst/UVIS spectra  
of the UV bright spot identified by~\citet{kin95} in the M31 nuclear region is similar 
to the one of A0-type stars. They agreed with the scenario proposed 
by~\citet{lau98} and~\citet{bro98} that the UV bright spot could 
represent a $\sim$200 Myr old stellar cluster.~\citet{lau12} further resolved  
a handful of UV-bright stars within the central 0.2\arcsec\ (0.8 pc),  
which were suggested to be young early-type main-sequence 
(MS) stars instead of old post early AGB (PEAGB) stars.~\citet{sag10} also 
found that the age of a single stellar population fitting the observed Lick indices 
decreased to 8 Gyr, inside the central 2\arcsec\ (8 pc), 
which was much younger than that beyond the nuclear region (12 Gyr); An extra young stellar 
population with age $<$ 600 Myr and mass fraction lower than 10\% could 
remove this age difference of old stellar population 
inside and beyond 2\arcsec . They suggested that a burst of 
star formation could have occurred within the nucleus $\sim$100 My ago. 

That raised the question 
%Then, the next question is 
whether this star formation activity and the corresponding 
young MS stars are constrained to only the central few arc seconds or not. 
\citet{don15} claimed that 
intermediate-age stellar population (300 Myr to 1 Gyr) could pervade the 
M31 bulge, while contributing only $\sim$1\% of the total stellar mass, which is also 
supported by~\citet{con16}. %, who used a pixel CMD method. 
 However, this question has not been completely 
settled,   
%However, a detailed study of the stellar population of the M31 
%bulge with the ultraviolet (UV) 
%optical color magnitude diagram (CMD) is still lacking, 
%due to the serve crowding. %At these bands, 
%intermediate-age stars are much brighter and bluer than the old evolved low 
%mass stars, compared to the near-infrared (NIR) band, so that it is easier to distinguish these stars.  
%It is known that the M31 nuclear are blue for many years
because %the disadvantage of 
previous studies  
with integrated spectra or photometry %from UV to NIR bands 
%is that the intensity is 
could be dominated by a few 
bright stars, such as AGB stars in the NIR band or numerous old evolved low 
mass stars, such as RGB stars in the optical band. Even in the UV, 
the integrated intensity could still suffer from contamination of the numerous extreme horizontal  
branch stars~\citep[][and references therein]{don15}.  
Meanwhile, limited by the total number of photometric 
data points, only one or two starburst activities could be 
used to fit the whole star formation history (SFH) in the M31 bulge~\citep{sag10,don15}. 
Therefore, %limited by the 
%available data points (photometric 
%observations) or wavelength coverage (spectroscopic observations), 
we can not have 
a clean scrutiny of the SFH in the M31 bulge. 
%The CMD fitting with 
%individual resolved stars does not have this problem and is the most accurate 
%method to disentangle stellar populations with different ages and metallicities, which 
%fall into various locations in the CMD. On the other hand, 
%while adaptive optics NIR images taken by ground-based $>$ 8m telescopes could indeed 
%achieve diffraction limit, such as 0.09\arcsec\ in~\citet{ols06}, their short color 
%range ($J$-$K$) hampers our attempt to disentangle stellar populations with different 
%ages and metallicities. Instead, because main-sequence stars is brighter 
%in UV and optical, the UV and optical CMD could be better to identify the 
%intermediate-age stellar population. 
On the other hand, CMD fitting technique uses all individual resolved stars and 
the result is not dominated by a few bright stars. Although the AO NIR images 
taken by ground-based $>$ 8m telescopes could indeed achieve diffraction limit, 
such as~\citet{ols06}, the limited NIR color 
range (e.g., $J$-$K$) hampers our attempt to disentangle stellar populations with different 
ages and metallicities. Instead, because MS stars are brighter 
in the UV and optical bands, the corresponding CMD could be better to identify the 
intermediate-age stellar populations.

%young 
%stars are recognized near the super massive black hole. On the other hand, 
%intermediate-age stars are suggested to penetrate in the central 4\arcsec , which 
%are suggested to be related to the secular evolution of the M31 bulge. Therefore, 
%carefully disentangling stars with different ages and metallicity in the M31 
%bulge is very critical for us to understand how it assembles. 

%However, our understanding of the stellar populations in the M31 bulge suffers 
%from the crowding issue. The color magnitude diagram fitting is the most accurate 
%method to disentangle stellar populations with different ages and metallicities, which 
%fall into different locations in the color magnitude diagram. However, limited by the 
%angular resolution of the ground-based telescope,~\citet{ols06} keep the record and 
%only study the regions with galactocentric radius $>$2\arcmin . Instead, most of 
%previous works based on the integrated stellar light, which is aggregation of the 
%luminosity of different stellar population and always dominated by 
%the bright stars, such as the AGB stars in the near-infrared and main-sequence turn-off 
%stars in the optical band. Therefore, limited by the available data points (photometric 
%observations) or wavelength coverage (spectroscopic observations), we can not have 
%a clean scrutiny of the star formation history in the M31 bulge. 

In this paper, we will utilize new deep UV and optical images 
taken by \hst\ to 
perform CMD fitting and further study the SFH of  
the M31 bulge. The angular resolution of \hst\ in the UV and optical bands is 
$\sim$0.05\arcsec , which is better than previous AO observation  
in the NIR band. On the other 
hand, \hst\ has a more 
stable PSF, a lower 
background and a larger field-of-view (FoV) than the ground-based 
AO observation. All of these properties 
above are important for us to 1) detect dim stars, 
2) reduce the photometric uncertainties in the 
crowded environments,   
 which are critical for the CMD fitting, and 3)
study spatial 
variations of different stellar populations.
 
The structure of this paper is as follows. In \S~\ref{s:data}, 
we present our observation. Then, we describe 
our methods to produce the Nyquist-sampled images, 
construct catalogues from detected sources in deconvolved images, 
perform artificial star tests,  
compare with the source catalogs from the pipeline of the 
Panchromatic Hubble 
Andromeda Treasury (PHAT) survey 
and discuss the foreground and background contamination 
in our catalog in \S~\ref{s:reduction}. 
%, we construct 
%deconvolved source catalogs. We also 
%compared the color magnitude diagram and luminosity function from 
%our deconvolved source catalog with those 
%from the Panchromatic Hubble Andromeda Treasury (PHAT) 
%pipeline. 
In \S~\ref{s:analysis}, we perform the fitting of 
the CMD %and luminosity function (LF) 
to 
disentangle stellar populations with different ages and metallicities in 
the M31 bulge and examine the blending effect on our CMD fitting. 
In \S~\ref{s:discussion}, we 
discuss the impact of the results on our understanding the assembly history 
of the M31 bulge. We will summarize our 
results in \S~\ref{s:summarize}.

\section{Observations}\label{s:data}
We utilize the dataset taken by the \hst/ACS WFC and HRC cameras. 
The FoV of these two cameras are 202\arcsec$\times$202\arcsec\ and 
29\arcsec$\times$26\arcsec , respectively. 
The former observation is from 
the PHAT survey~\citep[Program GO-12055,][]{dal12}, while the latter one 
is from Program GO-10571~\citep{lau12}. The detailed description of these 
two dataset are given 
in~\citet{dal12},~\citet{wil14} and~\citet{lau12}. Here, 
we only briefly summarize the key points, related to 
this work. 

\subsection{PHAT survey}\label{ss:survey}
%In the \hst\ ACS/WFC observations of the PHAT survey,  for each pointing,
%there were five and four dithered exposures at these two bands, 
%respectively. 
The PHAT survey used the F475W and F814W filters in the optical band to 
satisfy the high throughput and large wavelength leverage 
simultaneously. At these two bands, each pointing has five and 
four dithered exposures, respectively. The
basic information of the filters and exposure times of dithered images
is listed in Table~\ref{t:filters}. The short `guard' exposures (10 second 
in F475W and 15 second in F814W) were used to recover the photometry of bright 
stars, which saturated in the long exposures. 
Because the plate scale of the WFC camera (0.05\arcsec\ pixel$^{-1}$) is 
similar to the full
width half maxim (FWHM) of the point spread functions (PSFs)  
at these two bands, the PSF is undersampled. In order to overcome this problem, 
the survey used half pixel dithering in four and two 
dithered images with the longest exposure times at these two bands, respectively. 
In \S\ref{ss:nyquist}, only these exposures 
are used to produce the Nyquist-sampled
summed image. As a result, without the `guard' exposures, 
we cannot recover the photometry 
of stars with F475W $<$ 18.7 mag and F814W $<$ 18.6 mag.

\begin{figure*}[!thb]
  \centerline{
       \epsfig{figure=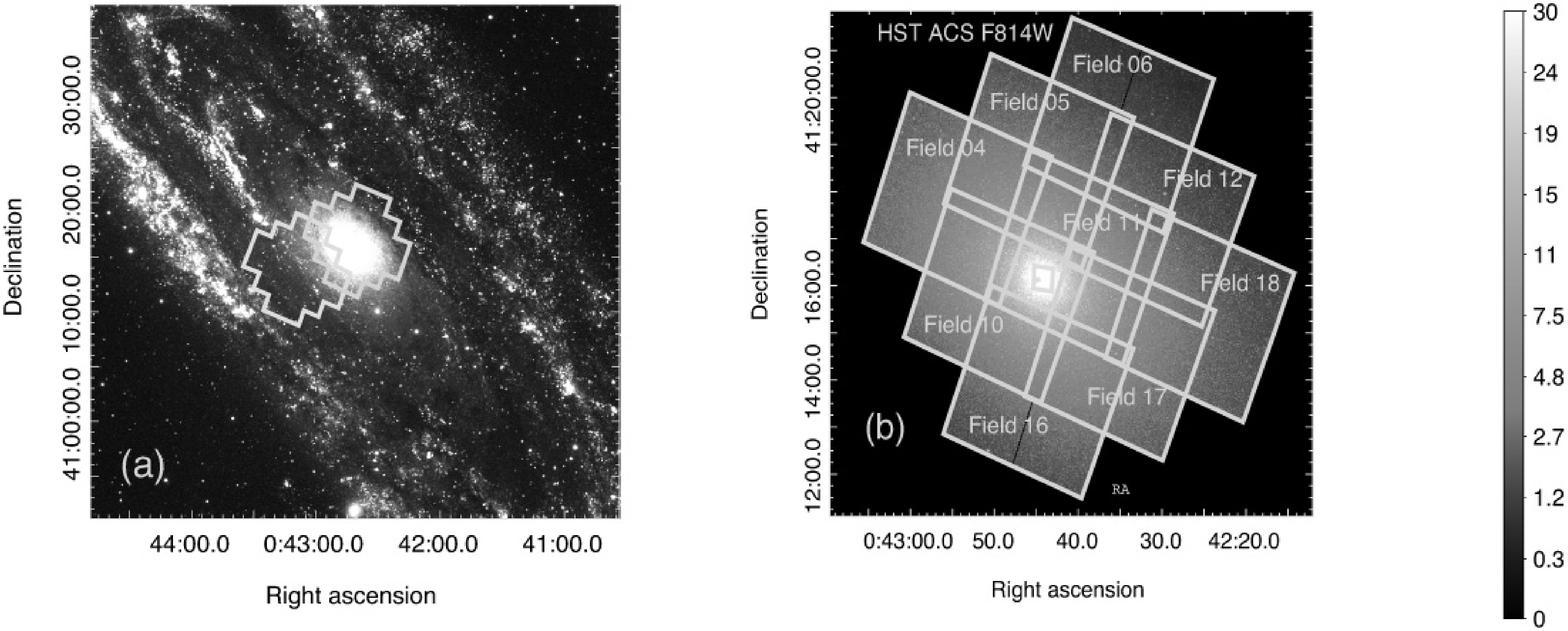,width=1.0\textwidth,angle=0}
       }
 \caption{(a): \galex\ composite mosaic of M31 (red: NUV, blue:
   FUV). The polygons show the boundary of Brick 1; Green
   and cyan polygons are halves of Brick 1 with and without half-pixel
   dithering strategy, respectively. (b):
   \hst\ \acs/WFC F814W mosaic. The units of the mosaic are 
   counts $s^{-1}$ $pixel^{-1}$. The green boxes and labels represent 
individual `Fields'. The cyan polygon shows the field-of-view of the 
ACS/HRC observation. }
 \label{f:field}
\end{figure*}

We used the data of Brick 1, one of the 23 Bricks in the PHAT
survey. Each Brick consisted of 3$\times$6 gird of pointings; each
pointing was dubbed `Field'~\citep{dal12}. We only concentrated on the 
half of Brick 1 (3$\times$3 grid of pointings), which includes M31* and the M31
nuclear bulge; unfortunately, the other half, did not 
operate with the half-pixel dithering strategy, which was required to produce the
Nyquist-sampled images below. The region used in
this study is illustrated in Fig.~\ref{f:field}a and the total sky area is $\sim$61 
arcmin$^2$ (870 pc$^2$). Besides the M31 bulge, the celestial northwest 
part of our FoV (see 
Fig.~\ref{f:field}a) covers a small portion of the 5 kpc star formation 
ring~\citep{lew15}.

%The data of these regions was collected from Dec 12th, 2010 to Dec 26th, 2010. 

%We also used the \hst\ WFC3/UVIS observations at the 
%F275W and F336W bands from the PHAT survey to 
%distinguish the main sequence stars and PAGB. Compared 
%to the F475W and F814W bands in the optical, the images of 
%these two UV bands have lower stellar number density and 
%suffer less confusion limit. Therefore, we do not produce the 
%Nyquist-sampled summed image for these two bands and only 
%employ the source catalog provided by the PHAT pipeline. 

\subsection{\hst\ ACS/HRC observation}\label{ss:acs}
\begin{figure*}[!thb]
  \centerline{
       \epsfig{figure=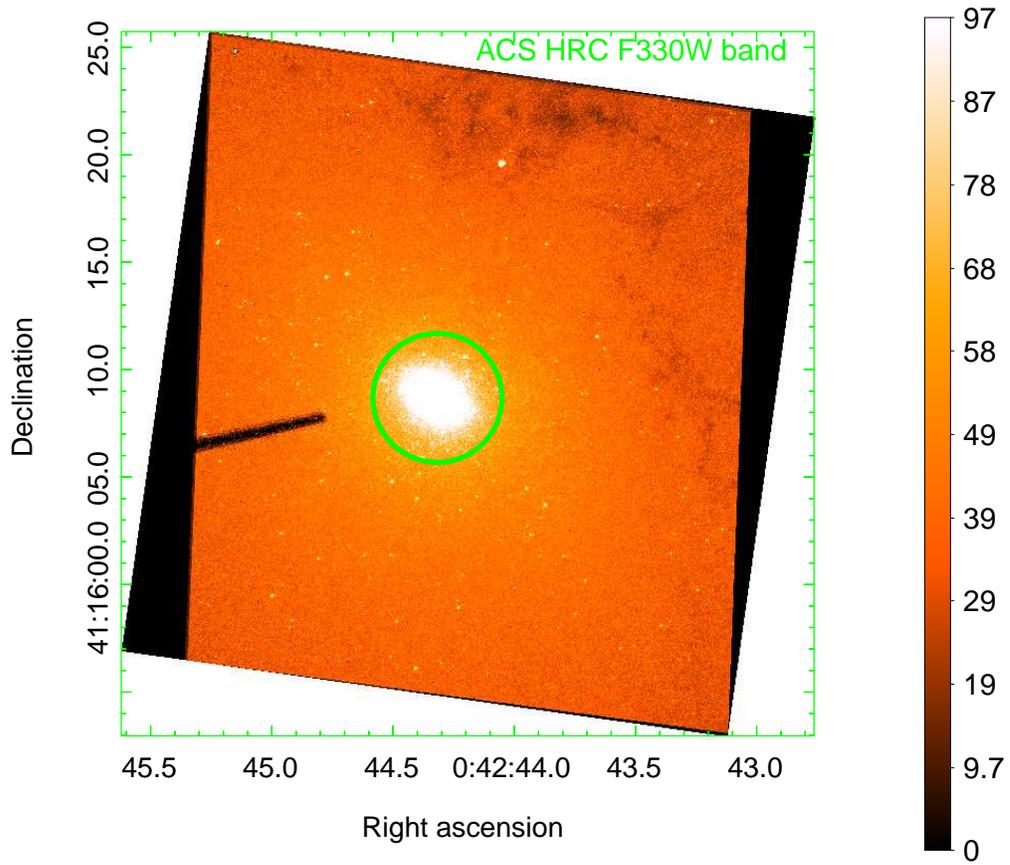,width=1.0\textwidth,angle=0}
       }
 \caption{The ACS/HRC F330W image of the M31 nucleus  
 ($\sim$29\arcsec$\times$26\arcsec ). The units of the image are 
   counts $s^{-1}$ $pixel^{-1}$. The radius of the green circle 
 is 3\arcsec .}
 \label{f:hrc_f330w}
\end{figure*}

Besides the PHAT dataset, 
the M31 nucleus was observed by the HRC camera at the 
F330W and F435W bands in the shorter wavelengths, with a sightly 
higher angular resolution (PSF FWHM $\sim$0.04\arcsec ). 
Half-pixel dithering pattern was used in the observation. 
The pixel scale of the HRC camera 
(0.028\arcsec$\times$0.025\arcsec ) is finer than that of the 
WFC camera by a factor of two. 
As a result, compared to the WFC observation, the PSF in the HRC 
images is better-sampled. %For example, Fig.~\ref{f:hrc_wfc_comp} 
%compares the HRC F435W and WFC F475W Nyquist-sampled 
%images derived in \S\ref{ss:nyquist} for a region in the celestial north of 
%the M31 nuclear region ($\sim$8\arcsec ). 
%The PSF in 
%the HRC observation is clearly sharper than that of the 
%WFC observation. 
Meanwhile, the two bands of the HRC observation  
in the near-UV provide us with a better chance to 
discover the intermediate-age stellar population, which is supposed to be 
brighter than the dominant old metal-rich population at this 
wavelength range. Therefore, the HRC observation is a good 
complement of the PHAT survey. Table~\ref{t:filters} gives the 
central wavelengths and widths of these two filters, as well as the 
exposure times. The saturation magnitudes of these two bands are 13.6 and 
14.8 mag. The final HRC F330W Nyquist-sampled 
image derived in \S\ref{ss:nyquist} 
is shown in Fig.~\ref{f:hrc_f330w}. 

\subsection{\spitzer/IRAC 3.6 $\mu$m observation}\label{ss:irac}
Besides the \hst/ACS observations, we used the \spitzer/IRAC 
3.6 $\mu$m observation of the M31 bulge~\citep{bar06} to constrain 
the spatial distribution 
of the stellar mass in the M31 bulge. As mentioned 
in \S\ref{s:analysis}, in the inner region, limited by our detection limit, the CMD cannot 
reach RGB stars older than 2 Gyr; As a result, we cannot 
constrain the stellar mass of the old stellar population, which is the 
major contribution of the total stellar mass. On the other hand, at 3.6 $\mu$m, 
the stellar light is less affected by the foreground extinction and 
the corresponding mass-to-light ratio is insensitive to the age and metallicity of underlying 
stellar population~\citep{mcg14,mei14}.~\citet{mcg14} and~\citet{mei14} give 
0.47 and 0.6 M$_{\odot}$/L$_{\odot}$, with Kroupa and Chabrier initial mass 
function (IMF)~\citep{kro01,cha03}, respectively. In this paper, 
we choose the value given in~\citet{mcg14} 
to convert the 3.6 $\mu$m intensity map into the stellar mass map, because 
we use Kroupa IMF in the CMD fitting below.  
%For example, in Starburst99 stellar synthesis 
%model~\citep{vaz05}, for a stellar population with solar metallicity, 
%the mass-to-light ratio only changes by 2.2 from 5 Gyr to 15 Gyr. 
%Considering that previous work~\citep{sag10,don15} 
%has suggested that the M31 bulge is dominated by the metal-rich 
%old stars, we used the median mass-to-light ratio for the stellar 
%population, which has Kroupa initial mass function (IMF), 
%super solar metallicity ($>$Z$_{\odot}$) and older than 10 Gyr, 
%to translate the 4.5 $\mu$m intensity map into the stellar mass map. 

%The final HRC F330W 
%image 
%is shown in Fig.~\ref{f:hrc_f330w} and 
%the comparison between the HRC F435W and WFC F475W  
%images derived in \S\ref{ss:nyquist} is given in 
%Fig.~\ref{f:hrc_wfc_comp}. 
%  As a result, with the method described below, we can still achieve Nyquist 
%sampling images. 

\section{Data Reduction}\label{s:reduction}
\subsection{Nyquist-sampled images}\label{ss:nyquist}
\citet{lau12} have analyzed the same HRC dataset 
with a Fourier method, which was 
originally presented in~\citet{lau99}. 
We employed the same technique  
to reduce the WFC observation and produce Nyquist-sampled 
images for individual pointings. 
Unlike the widely used `AstroDrizzle', the technique of~\citet{lau99} 
does not cause any smoothing and degradation of the PSF. The detailed 
description of the procedure is given in~\citet{lau99} and~\citet{lau12}. 
We simply describe some key points below. 

We constructed Nyquist-sampled images from calibrated dithered 
exposures kindly provided by the PHAT pipeline team. These exposures 
were bad pixels removed, bias and dark subtracted, as well as 
flat-fielding corrected. %, with the \hst\ pipeline, {\tt OPUS}
% version 2010\_4 and {\tt CALACS} version 5.1.1. 
Charge transfer efficiency which the WFC camera suffers was amended. 
All the dithered exposures of the same filter and pointing were aligned. 
We first rescaled the calibrated images at the F475W and F814W bands 
 in units of counts to have 
360 and 800 seconds respectively. They were then interlaced 
into a roughly Nyquist-sampled image with a square-spiral pattern. %, after correcting for 
%t
The pixels with abnormal count
rates were selected out and replaced by the interpolation from 
nearby pixels. Second, we used the Fourier algorithm of~\citet{lau99} to
combine the repaired dithered exposures into a well-sampled Nyquist
images. This algorithm sums the Fourier transforms of the dithered 
exposures, suppresses the high-frequency power
and transforms the result back into imaging space, but with half of the original
WFC pixel size, i.e. 0.025\arcsec/pixel. Third, using the calibration files 
from the STScI, we corrected for the distortion and multiplied the pixel area map 
to get the final Nyquist-sampled images at different pointings and filters. 

For each Nyquist-sampled image, we also produced a sky background and derived a  
noise value for source detection below. Individual Nyquist-sampled images were 
smoothed with a median box, the size of which is 250$\times$250 pixel 
(6.25\arcsec$\times$6.25\arcsec ) and 100$\times$100 pixel 
(1.14\arcsec$\times$1.14\arcsec ) for the WFC and HRC observations, respectively, to 
be their sky background images. The histogram of the pixel values of individual 
Nyquist-sampled images includes a Gaussian function plus a tail toward the bright end;  
The former one represents the fluctuation introduced by the bias and sky background 
and the latter one is contributed by bright stars. We derived the standard deviation of the 
Gaussian function and used it to be the noise of each Nyquist-sampled image. 

Then, at each band, we merged these Nyquist-sampled images  together to the final mosaic. We
employed the method detailed in~\citet{don11} to derive the relative
astrometry and bias offset among different pointings, by using their overlap regions. 
Then, we used bright isolated stars to align our images to the
absolute astrometry system defined by the 2MASS
catalog~\citep{skr06}. The F814W mosaic of our data is shown in Fig.~\ref{f:field}b with 
the boundaries and labels of the `Fields'. % and the FoV of the HRC 
%observations. 
 
%\subsection{Photometry}\label{ss:photometry}
%In \S\ref{ss:decon}, we perform the source detection and 
%extract the deconvolution photometry 
%on the Nyquist images for each pointing of the ACS WFC and HRC observations 
%at the four bands, respectively. Hereafter, when we mention `two bands', 
%we mean F330W and F435W 
%for HRC observations or F475W and F814W for WFC observations, 
%respectively. Then, we compare our deconvolution photometry with the
%{\tt DOLPHOT} photometry from PHAT pipeline team 
%in \S\ref{ss:compare} 
%and demonstrate the advantage of
%the deconvolution photometry in the crowded environment. PHAT
%pipeline team provides two versions of source catalogs (`st' and `gst') in one pointing
%with different criteria (see~\citealt{dal12,wil14}). The latter one is used more strict 
%criteria. Therefore, in \S\ref{ss:compare} %In the `st' catalog,
%sources remains, if either they have signal-to-noise ratio (SNR) $>$4 in
%both filters or have SNR $>$4 and the square of
%sharpness $<$0.1 in at least one band. On the other hand, the `gst' catalog just
%includes the high quality sources, i.e. their SNR $>$4 in both filters. 
%We select the `st' catalog for the comparison. % in the following
%sections. 

\subsection{Deconvolution photometry}\label{ss:decon}
We first empirically constructed the PSFs from bright isolated sources. 
Considering the spatial variation
of PSFs throughout the large FoV of the WFC camera, 
we derived the PSFs for the four
quadrants of the two chips, respectively. In the M31
bulge, limited by the crowding, very few isolated stars in individual quadrants of one pointing 
are bright enough to produce the PSFs. Therefore, we
aggregated isolated bright stars ($m_{F475W}$$<$20, $m_{F814W}$$<$19.5) 
in the same quadrant of the nine
pointings, considering that these images were taken 
during the same period (around half month) and the \hst\ PSF was very stable. We removed 
faint point sources near these bright stars, normalized and
median-averaged their images to obtain the final PSFs at different quadrants in
the two chips, respectively
(i.e. 4$\times$2=8 PSFs for each band). For the HRC observation, 
considering its small FoV, only one PSF 
was produced for the F330W and F435W observations, respectively. Since numerous 
RGB stars are dimmer in these short wavelengths, the surface number density 
is low and there are still many isolated bright sources which 
could be used to produce the PSFs.

Then we used the empirical PSF derived above to deconvolve the Nyquist-sampled 
images using the Lucy-Richardson algorithm. This algorithm worked 
iteratively to reduce the wing of the PSF and enhance its core, simultaneously. 
Therefore, in the deconvolved images,  individual stars
became delta functions. We found that 320, 160 and 640 iterations were enough 
for the WFC, HRC F330W and F435W  Nyquist-sampled images, respectively.

We detected sources and extracted their photometry from the deconvolved
images. First, we selected the pixels in the deconvolved image with values larger than the
background image by three times of the noise (the background images and the noises 
were produced in \S\ref{ss:nyquist}). 
We refer to these pixels as `source' pixels. 
We sorted them
according to their intensities. Second, starting with the brightest one, we considered 
that the adjacent 3$\times$3 pixels belong to the same source and then produced a source list. %If for
%some pixels, the adjacent 3$\times$3 regions include the pixels, which
%are already considered to be belonged to another brighter pixels, we
%jump this pixel. 
Third, we merged the source catalog of the two bands (hereafter, `two bands' 
refer to F475W and F814W for the WFC camera or F330W and F435W for the 
HRC camera)
 into a master catalog. Two sources at these two bands were considered 
as the same one, if their distance was less than 2 pixels, i.e. 0.05\arcsec (0.025\arcsec ),  
in the WFC (HRC) observation. In this master 
catalog, we also kept the
sources detected in only one band. Fourth, the
intensity of each source in the master catalog at each band was the
difference of the sum of values of its `source' pixels in the Nyquist-sampled image and the
background image. Fifth, we calculated the signal-to-noise ratio of
each source as the ratio of its intensity and the square root of the
sum of values of its 
`source' pixels in the Nyquist-sampled image (which is equal to intensity plus
background). Sixth, we translated the intensity
into magnitude with the zero magnitudes: 22.909, 25.188, 26.157 and 25.523 mags for the F330W, F435W, 
F475W and F814W bands, 
respectively\footnote{http://www.stsci.edu/hst/acs/analysis/zeropoints}. Seventh, 
for the WFC observation, 
we compared our source list of each position with that from the PHAT pipeline
 and derived the mean difference is only $\sim$0.05 mag. We subtracted this value from the
source catalog of each pointing. In our final catalog, we only
included the sources with signal-to-noise (S/N) ratio $>$3 in both two 
filters of the WFC and HRC observations. Meanwhile, because we did not use the guard exposures in the 
WFC observations when
producing the Nyquist-sampled images, we could not recover the photometry for the
bright sources. We identified a
saturated source, if any of its adjacent 11$\times$11 pixels had   
count rate $>$109 (95) electrons/s at F475W(F814W) bands, according to the 
ACS Instrument handbook. These stars were removed 
away from our discussion below. Eighth, for the WFC observations 
we merged the source catalogs of the nine pointings
into the final source catalog for the CMD fitting in
\S\ref{s:analysis}. 

\subsection{Artificial star test}\label{ss:artificial}
We generated artificial stars with -5$<$F475W-F814W$<$10 and 
13$<$F814W$<$29 (-4$<$F330W-F435W$<$4 and 18$<$F435W$<$28), 
added them back into the original WFC (HRC) 
images and re-performed the
source detection. Through comparing the input and output magnitudes,
we determined the recovered fraction (i.e., the fraction of input 
artificial stars, which were detected) and uncertainties as a
function of magnitude, color and local surface brightness. The dimmest  
magnitudes 
of the artificial stars were chosen to be one magnitude 
lower than the cut-off magnitudes of our 
source catalogs to examine the potential 
Eddington-bias. 

We used the Padova stellar 
evolutionary tracks~\citep[version: PARSEC v1.2S +COLIBRI PR16,][]{mar13,mar17,ros16} 
to generate the artificial catalog, which  
consists of two groups, both of which have similar numbers of stars: 
1) a stellar population with continuous star formation, as well as various metallicities 
(from 1./200 to 2.5 $Z_{\odot}$)
and ages (from 4 Myr to 13 Gyr). This component reflects the color and magnitude distributions 
of all potential stars. 2) a stellar population with old age ($>$ 5 Gyr) and high
metallicity ($>$$Z_{\odot}$). These red and 
dim stars dominate the M31 bulge and are near or below the detection limit of our
observations (see below). As a result, only very few of these input stars in the 
previous group have been 
detected and the derived recovered fraction at this color and magnitude ranges 
in the CMD suffers from 
large statistic uncertainty. 
%the recovered fraction of these stars are very 
%low, which is hard to be constrained by the previous group. 
Therefore, we added 
a new old and metal-rich component in the artificial catalog to enhance 
the number of recovered stars near the detection limit. 
% and better constrain the 
%recovered fraction at this part of the CMD.

For the WFC dataset, we manufactured ten observations 
at each pointing and band for the two groups above. In
each manufactured observation, we injected $\sim$1$\times10^5$ artificial 
sources with the PSF constructed in \S\ref{ss:decon} 
 into the original Nyquist-sampled image of the WFC observation, which was 
$\sim$5\% of the stars found in each pointing. For the HRC observation, 
we produced 10$^3$ artificial observations, each including 350 artificial sources. 
We ran the deconvolution routine and performed the source
detection. We recorded the positions,
magnitudes and the galactocentric radii ($R$) of individual output artificial stars.    

%Considering that the completeness and photometric 
%uncertainties are determined by 
%the local surface brightness, we divide all the
%artificial stars in the WFC pointings into nine annuli, centered 
%in the M31 nucleus,  
%considering the fact that the surface brightness 
%is axisymmetric in the M31 bulge  
%and changes quickly along the 
%galactocentric radius. 

\begin{figure*}[!thb]
  \centerline{
       \epsfig{figure=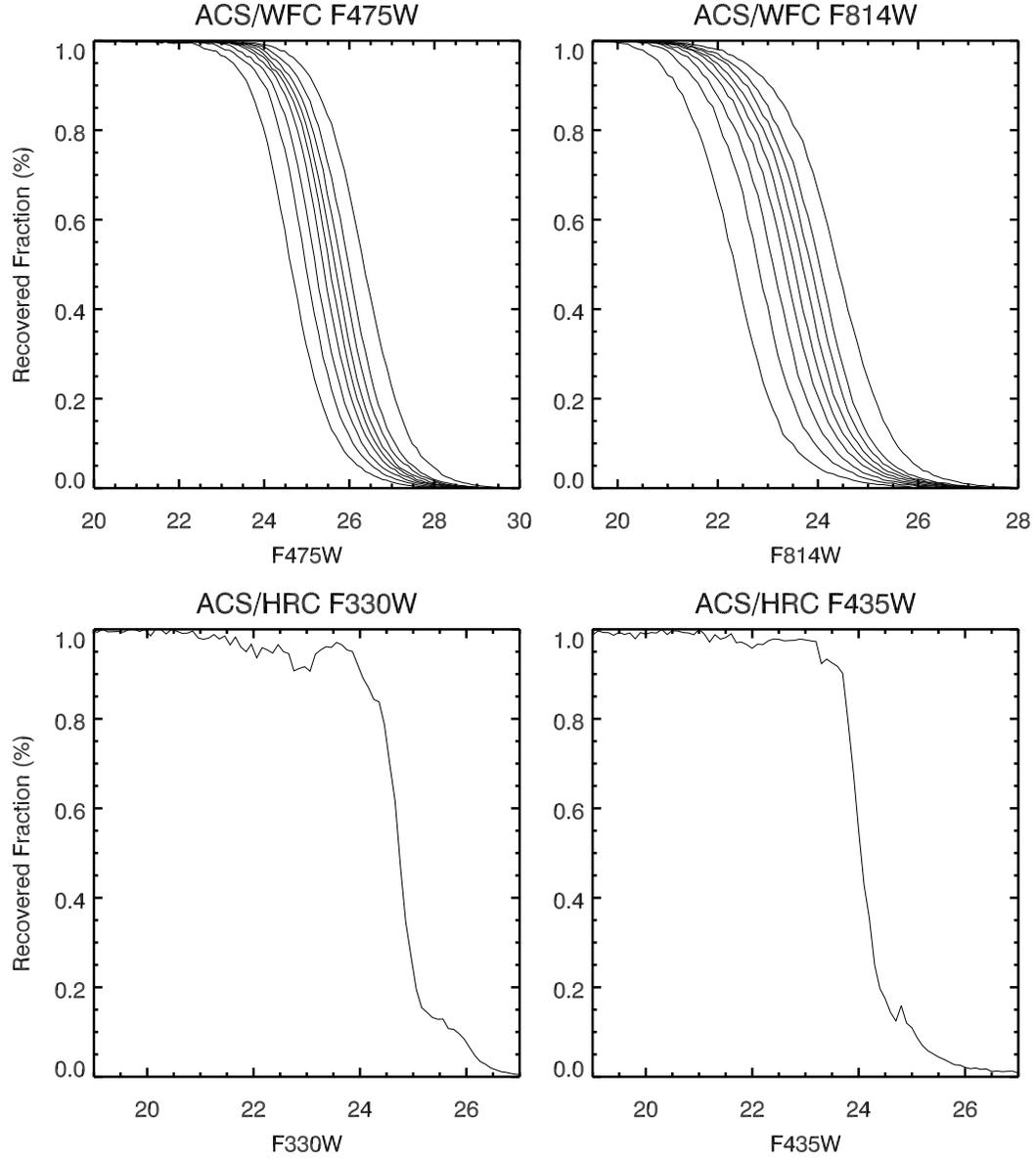,width=1.0\textwidth,angle=0}
       }
 \caption{The recovered fraction as a function of input magnitude derived 
 from the artificial star tests for the WFC (top panels) and HRC (bottom panels) 
 observations. In the top panel, the solid lines from left to right represent 
 the annuli from innermost to outermost with decreasing surface densities.} 
 \label{f:det_limit}
\end{figure*}

Besides the input magnitude, 
the recovered fraction and photometric uncertainties are determined by 
local surface brightness too. In the M31 bulge, the surface brightness 
is axisymmetric with respect to the M31 nucleus~\citep{bea07,don14,don15} 
and changes quickly along $R$. 
%galactocentric radius. 
As a result, due to the large FoV, even in the same WFC pointing, 
the recovered fraction  
and photometric uncertainties could change at different locations 
of the camera. Thus, 
instead, %of using the artificial stars to constrain the completeness and 
%uncertainties in individual 
%fields, 
we equally divided the
artificial stars in all the pointings, according to their $R$, 
%galactocentric radius, 
into nine annuli. The inner and outer $R$
%galactocentric radii 
of these nine
annuli are given in Table~\ref{t:annu}.  When $R$ %galactocentric radius 
increases, the stellar number
surface density decreases and our source catalog is less affected by
the blending, especially for faint sources. Thus, the recovered fraction increases.  
Fig.~\ref{f:det_limit} shows the recovered fraction
as a function of the input magnitude at the nine different radii. 
The 50\% 
detection limit of the F475W and F814W bands decreases from 26.3 and 24.4 mag 
for the outermost ring by $\sim$2 mag to 24.6 and 22.3 mag in the innermost ring. 
For the HRC observation, 
due to the small FoV, we did not divide it into finer parts. The 50\% 
detection limit of the F330W and F435W bands are 24.6 and 23.9 mag, respectively. 
The recovered fractions of these two bands are also given in Fig.~\ref{f:det_limit}. 
%The completeness fraction is anti-correlated with the input
%magnitude. The dimmer sources are hard to be detected, easily hide
%in the wing of the bright sources or blend to be a brighter
%source. 

\begin{figure*}[!thb]
  \centerline{
       \epsfig{figure=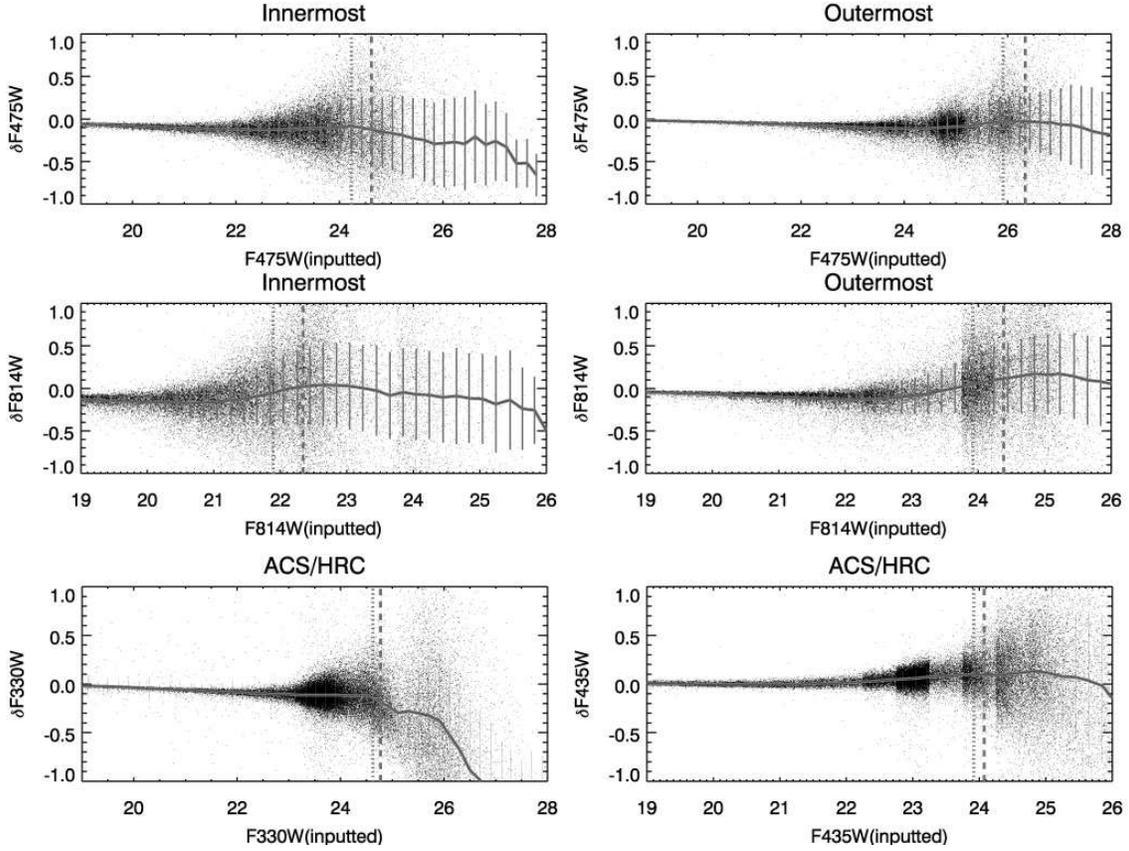,width=1.0\textwidth,angle=0}
       }
 \caption{The difference between the output and input magnitudes at the 
   F475W (top row) and F814W (middle row) bands for the 
   innermost and outermost
   annuli of the WFC observation, as well as the HRC observation (bottom row). 
   The red lines and bars represent the mean and standard
   deviation of the differences at individual input magnitude bin. The red dotted and 
   dashed lines represent the 70\% and 50\% completeness limits. }
 \label{f:det_err}
\end{figure*}

Fig.~\ref{f:det_err} shows the difference between the output and
input magnitudes of the artificial
star tests for the innermost and outermost radii of the WFC observation, as well as the HRC observation, 
as a function of the input magnitudes. The deviation for the bright stars is
%negligible. 
small.
Instead, it increases towards faint magnitudes. Due to 
blending, the output magnitude is averagely smaller than the input one. For 
bright stars, the deviation is larger at the F814W band than the F475W, F435W and F330W 
bands, indicating the crowding issue is more significant at the long wavelength. %There is gaps of stellar number density at F475W$\sim$25.5 and F814W$\sim$23.5, respectively. 
%That is caused 

%because of the enhancement of red clump stars in our artificial catalogs. %The
%completeness fraction decreases with the increase of the
%magnitude. However, due to the jump of total number of stars at
%F475W$\sim$26 and F814W$\sim$24, we see the enhancement of stars
%around these magnitudes. 

\subsection{Comparison with the DOLPHOT photometry}\label{ss:compare}
We compare the performance of our deconvolution method and the {\tt DOLPHOT} algorithm 
used by the PHAT pipeline~\citep{dal12} in the extremely crowded environment, like the M31 bulge, 
through the real observation and artificial star clusters. 

\subsubsection{Real observation}\label{ss:real}
\begin{figure*}[!thb]
  \centerline{
       \epsfig{figure=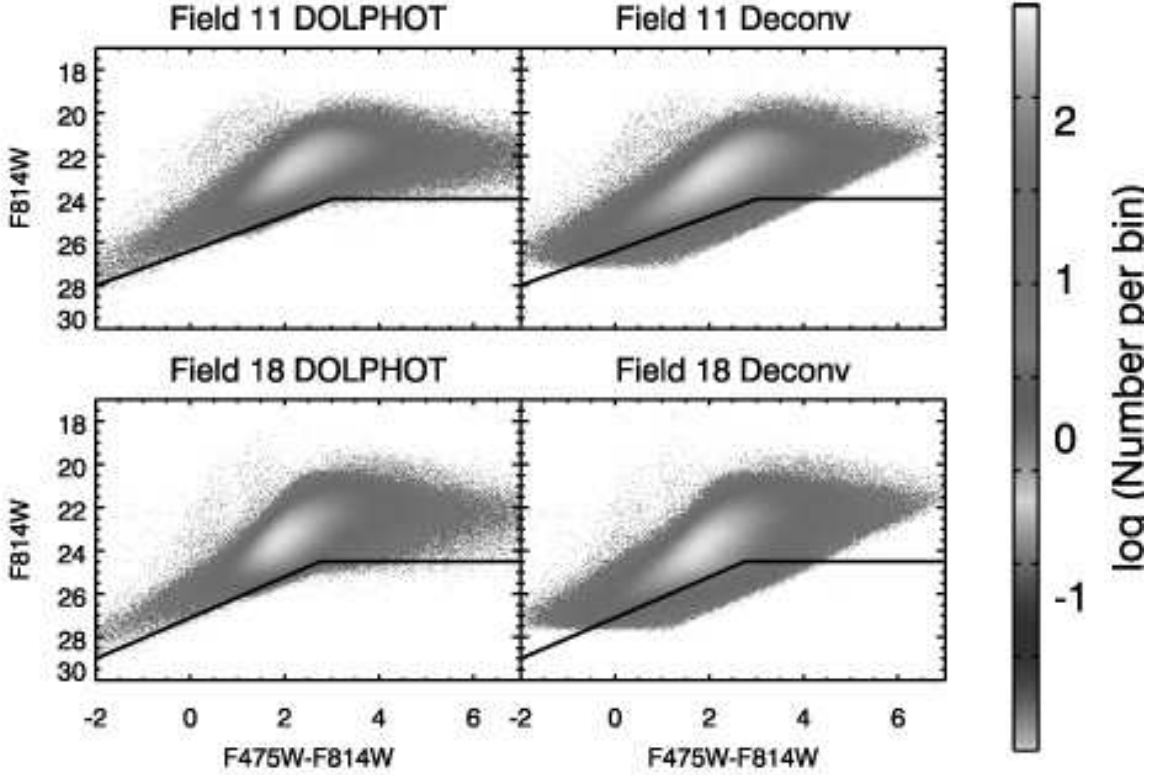,width=1.0\textwidth,angle=0}
       }
 \caption{The Hess diagrams for Field 11 (top panel) and Field 18
   (bottom panel), with the {\tt DOLPHOT} `st' catalogs from PHAT pipeline
   team (left column) and our deconvolution catalogs (right
   column). These two `fields' have the highest and lowest star 
   number surface densities in our study. Compared
   to Field 11, the detection limit of Field 18 has increased by $\sim$ 1
   mag at the F814W band. The black solid lines in the Hess diagram of these two `Fields' 
roughly mark the
   detection limit of the {\tt DOLPHOT} `st' catalog. Our
   deconvolution method finds more dim and red stars in the low right
 part of the Hess diagram. The sharp cut of the Hess diagram of our
 deconvolution catalogs at the dim and red side is due to the S/N cut at both bands. 
Compared to our deconvolution catalog, the {\tt DOLPHOT} `st' catalog 
has extra sources at 1) F475W-F814W$>$6 and
2)F475W-F814W$<$0 and F814W$>$27, which are due to their S/N cut at only one band. }
 \label{f:dol_dec_com}
\end{figure*}

\begin{figure*}[!thb]
  \centerline{
       \epsfig{figure=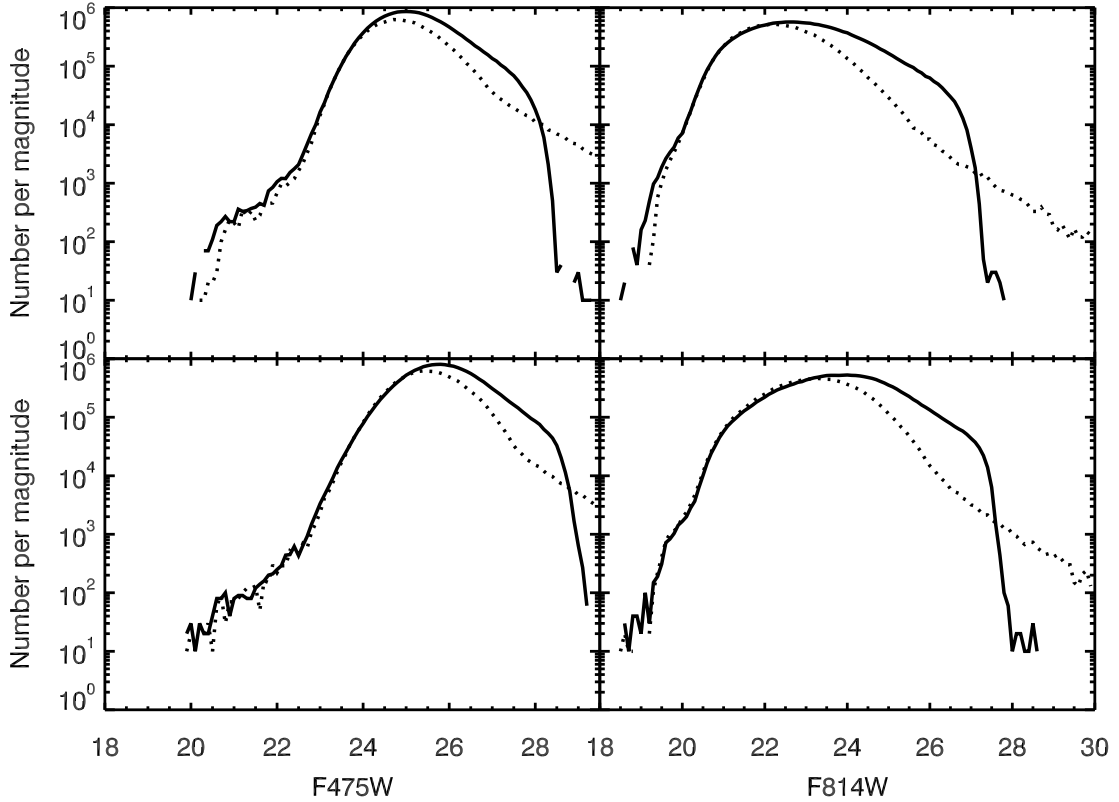,width=1.0\textwidth,angle=0}
       }
 \caption{The LF for Field 11 (top panel) and Field 18
   (bottom panel), at F475W (left column) and F814W (right column)
   with the {\tt DOLPHOT} `st' catalogs from PHAT pipeline
   team (dashed lines) and our deconvolution catalogs (solid 
   lines). Our method detects more dimmer sources. The shape break of
   the LF from our deconvolution catalog is due to
   the S/N cut at both bands. Instead, the long tail in the
   LF from the {\tt DOLPHOT} `st' catalog is also due to
   their S/N cut at only one band.}
 \label{f:dol_dec_lum}
\end{figure*}

Figs.~\ref{f:dol_dec_com} and~\ref{f:dol_dec_lum} are the comparisons 
of the CMD and luminosity function (LF) between 
the {\tt DOLPHOT} `st'
catalogs\footnote{The PHAT pipeline provides two catalogs for each `Field': `gst' and `st'. The 
former uses stricter quality control. As a result, the incompleteness is high in the crowded regions. 
Therefore, we used the `st' catalog for comparison.} and our ones at two `Field's: Field 11 and
Field 18. These two pointings have the highest and lowest number surface
densities in our study. As a result, the detection limit at the F814W band in 
Field 18 is $\sim$1 mag below that in Field 11. The CMD and LF derived 
from the {\tt DOLPHOT} and
deconvolution method are overall very similar. 
However, compared to the {\tt DOLPHOT} catalogs, our deconvolution
method can detect more dim and red stars in both fields; for example, 
our deconvolution method detects 
roughly four times more stars below the detection limit of 
the F814W band than {\tt DOLPHOT}. Because we required the
sources to have S/N$>$3 at both F475W and F814W bands, 
our LF has a clear cut at the dim sides of these two bands.  

\subsubsection{Artificial star clusters}\label{ss:artificial_cluster}
We inserted 28 artificial stellar clusters 
into original observation of Brick 17 and Field 06.  %(see Fig.~\ref{f:fake_source}). 
The salient parameters of these star clusters are
listed in Table~\ref{t:fake}. Compared to the real observation, the advantage of
this strategy is that we know the exact input magnitudes and positions of 
artificial stars. These clusters span a large range in
the total mass and compactness, from the very massive and compact ones,
such as, ID 15 and ID 19, to small and loose ones, for example,
ID 8. Thus, we could compare the 
performance of {\tt DOLPHOT} and deconvolution
photometry at different environments. 

\begin{figure*}[!thb]
  \centerline{
       \epsfig{figure=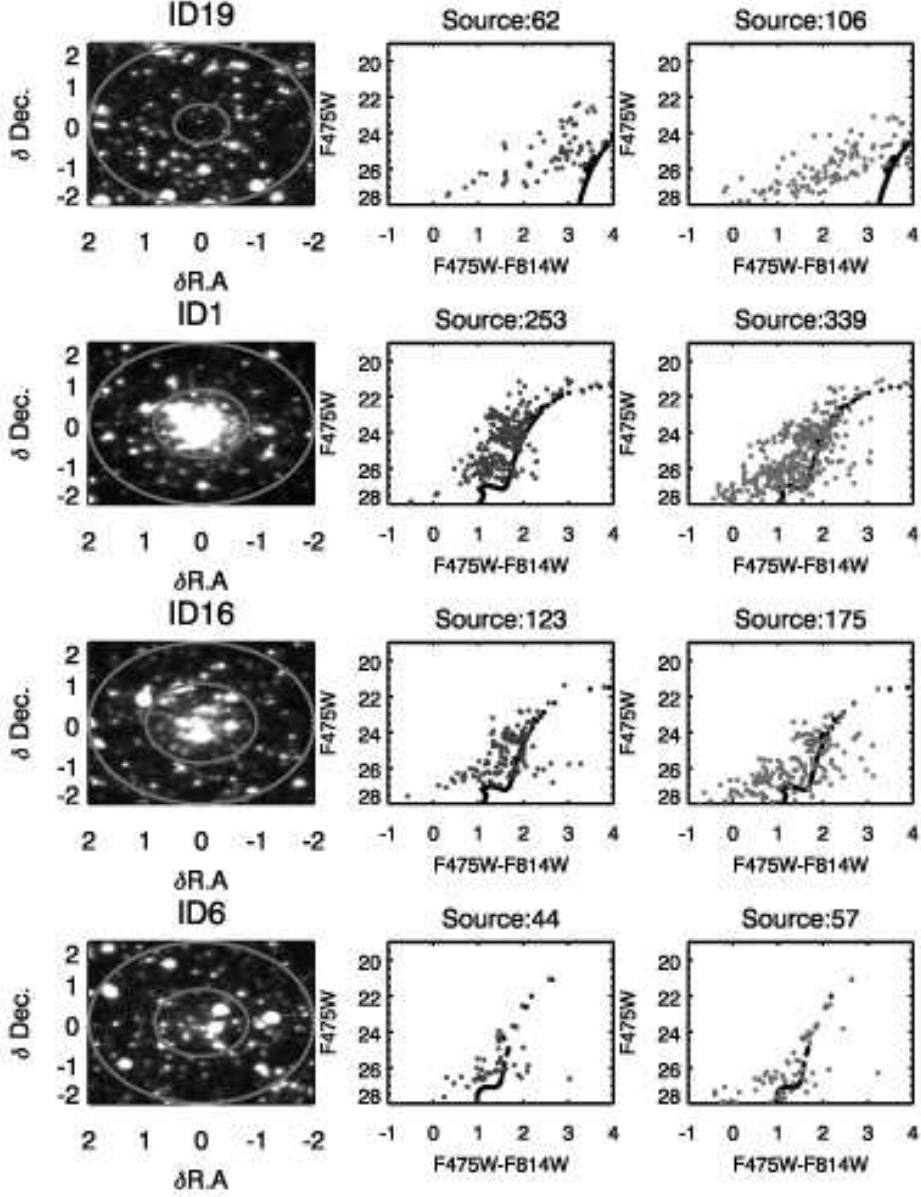,width=1.0\textwidth,angle=0}
       }
 \caption{Comparison of the CMD of {\tt DOLPHOT} (Middle
   column) and deconvolved catalog (Right column) in four selected
   artificial star clusters. Left column is the WFC F814W images of
   individual star clusters. The radii of the red and magenta circles are
   the effective radii and 2''. We extract the CMD 
   in the middle and right columns from the magenta circles. In the
   middle and right columns, the black circles are the input artificial 
   sources, while the blue and magenta circles are the sources detected
   by {\tt DOLPHOT} and deconvolution methods, respectively.}
 \label{f:fake_cluster_stack}
\end{figure*}

We used both our deconvolution method and {\tt DOLPHOT}
algorithm with the parameters given in~\citet{dal12} 
to extract the photometry. Fig.~\ref{f:fake_cluster_stack} shows the WFC 
F814W images, the {\tt DOLPHOT} and the deconvolved CMDs 
in the central 2\arcsec , with the field sources removed, 
of four representative star clusters. These clusters
were sorted according to their mass densities. ID 19 is the most compact
artificial star cluster. However, the large foreground extinction attenuates the 
brightness of individual stars. Therefore, most of the sources 
detected by the {\tt DOLPHOT} and 
deconvolved method should be the blending of several faint
sources and become brighter than the input magnitudes. 
On the other hand, ID 1, has much less
attenuation. Thus, we can see an apparent compact star cluster in the
F814W image. Interestingly, the {\tt
  DOLPHOT} CMD shifts to the brighter side of the input
CMD, while the deconvolved and input CMDs match 
better. The similar phenomena is found in the ID 16,
which is less compact than ID 1, by a factor of $\sim$3. In conflict, for
ID 6, the mass density of which is smaller than that of ID 1 by an order, 
both the CMDs derived by the {\tt DOLPHOT} and deconvolution method are
similar to the input one.

\begin{figure*}[!thb]
  \centerline{
       \epsfig{figure=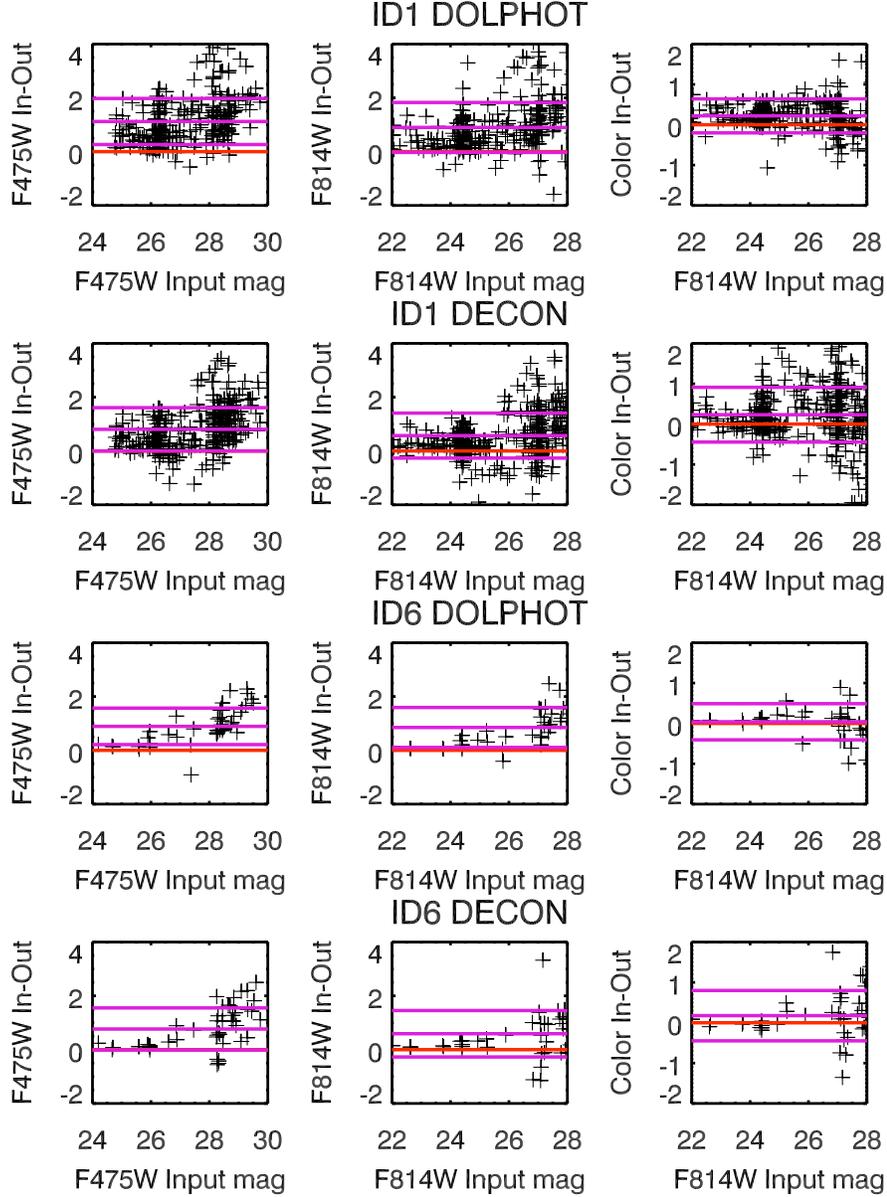,width=1.0\textwidth,angle=0}
       }
 \caption{The difference between the input and output parameters;
   magnitude and color, as a function of the input magnitude for two
  artificial star clusters.The red lines represent the same input and
   output parameters. The three cyan lines are the mean and $\pm$ one
   standard deviation of the difference between the input and output
   parameter. In some panels, the cyan lines overlays on the red lines. 
   The data in the first and third rows are from the
 {\tt DOLPHOT} catalog, while those in the second and fourth rows are
   from the deconvolution catalog.}
 \label{f:fake_cluster_mag}
\end{figure*}

We further compared the input and output magnitudes 
and colors of individual stars to understand the difference between the catalogs 
 derived by the {\tt DOLPHOT} and
deconvolution methods in Fig.~\ref{f:fake_cluster_mag}. 
We chose two artificial star clusters, ID 1 and ID 6. %, 
%which have different compactness. In both clusters, 
The output magnitudes of dim stars are smaller than the input ones, due to the
blending. %At the bright side, compared to ID 1, the output magnitude of
%stars in ID 6 with F475W$<$26 and F814W$<$25 are the same as the input
%one. On the other hand, due to the severe crowding, in ID 1, the blending
%already increases the brightness for these bright stars. 
Unlike the
magnitude, the input and output colors are very similar ($\leq$0.3 mag). 
That's because
the blending effect at both the F475W and F814W bands cancel each other. 
In ID 1, the deconvolved method performed better in the
photometry than the {\tt DOLPHOT} method, in terms of the smaller difference
between input and output magnitudes. For example, the mean and standard deviation of
the F814W magnitude are 0.9$\pm$0.93 mag for {\tt DOLPHOT} and 0.57$\pm$0.83 mag for
deconvolution. This explains why the output CMD from the {\tt DOLPHOT} 
algorithm is brighter than that from the 
deconvolution method. 

\begin{figure*}[!thb]
  \centerline{
       \epsfig{figure=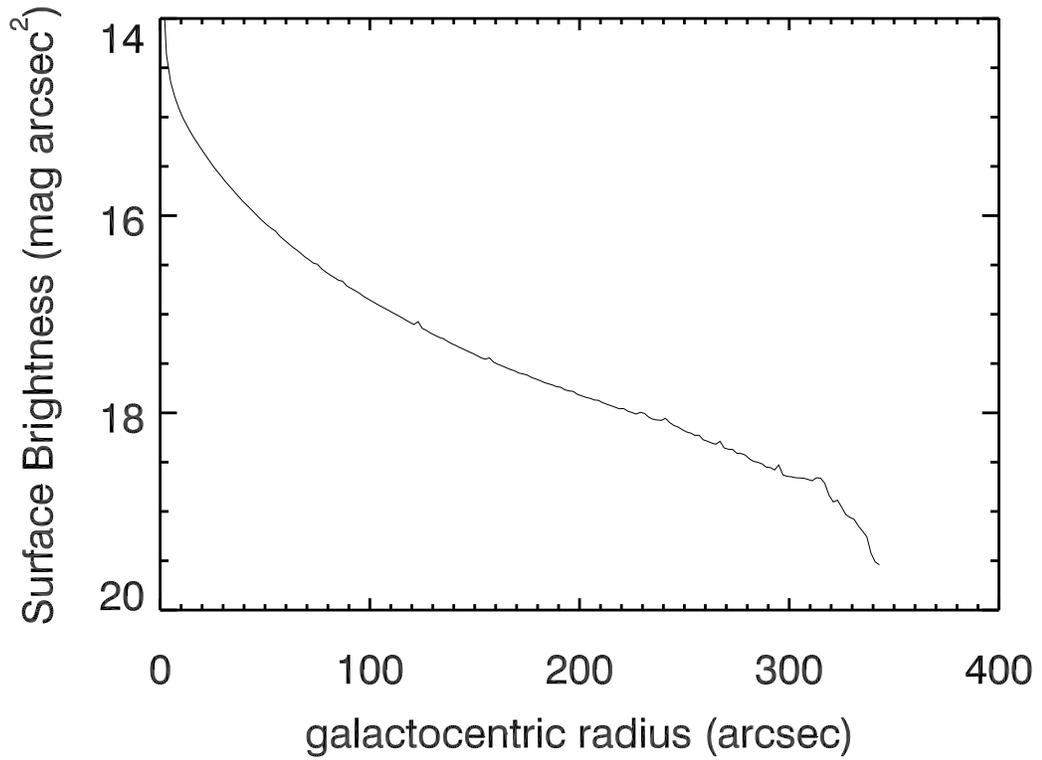,width=1.0\textwidth,angle=0}
       }
 \caption{The surface brightness (mag arcsec$^{-2}$) at the F814W 
band along the galactocentric radius ($R$, in units of arc second). The region inside the 
230\arcsec\ has brighter surface brightness than that of artificial cluster ID 1, which 
indicates that the deconvolution catalog should be preferred in our study.}
 \label{f:f814w_sb}
\end{figure*}

Fig.~\ref{f:f814w_sb} shows the surface brightness (mag arcsec$^{-2}$) at the F814W 
band as a decreasing function of $R$ (in units of arc second) in our FoV. 
The region inside the central 
230\arcsec\ is brighter than ID 1 (18 mag arcsec$^{-2}$). Therefore, 
the deconvolution method could provide us with a better catalog 
to study of stellar population in the M31 bulge. 

%which indicates that the deconvolved 
%catalog should be preferred in the study of stellar population in the M31 bulge through the CMD 
%or LF. 

\subsection{Foreground and background contamination}
Because the galactic latitude of M31 is low, 
our detected sources should 
include foreground Galactic dwarf stars, which have 
1$<$F475W-F814W$<$4. However, they account for a very tiny fraction of 
our observed stars. For example, we used the `Trilegal' code~\citep{gir12} to  
simulate the stellar population of the Milky Way in the FoV of our WFC 
observation. The code 
outputs 117 stars and only 113 of them are brighter than the 50\% completeness 
limit at both the F475W and F814W bands of the outermost \#9 annulus. The 
HRC observation should includes even much less contribution from 
the foreground Galactic dwarf stars, because of the smaller FoV and the 
faintness of dwarf stars at the UV bands. 

%Background galaxies are normally excluded from our catalog 
%by the sharpness cut.~
\citet{dal09} suggested the number density 
of background galaxies in the optical bands to be 
60 $arcmin^{-2}$~\citep[see also][]{wil14}. Therefore, we only expect less than 
one background galaxy in our FoV.

\section{Analysis and Results}\label{s:analysis}
In \S\ref{ss:cmd}, we first qualitatively analyzed different components in 
the CMD derived from our deconvolution catalog. Then we 
quantitatively fit the CMD to get the SFH in the M31 bulge in 
\S\ref{ss:cmd_fit}. After that, in \S\ref{ss:blending}, 
we examine the blending effect on our result based on the CMD fitting.  

\subsection{CMD}\label{ss:cmd}
\begin{figure*}[!thb]
  \centerline{
       \epsfig{figure=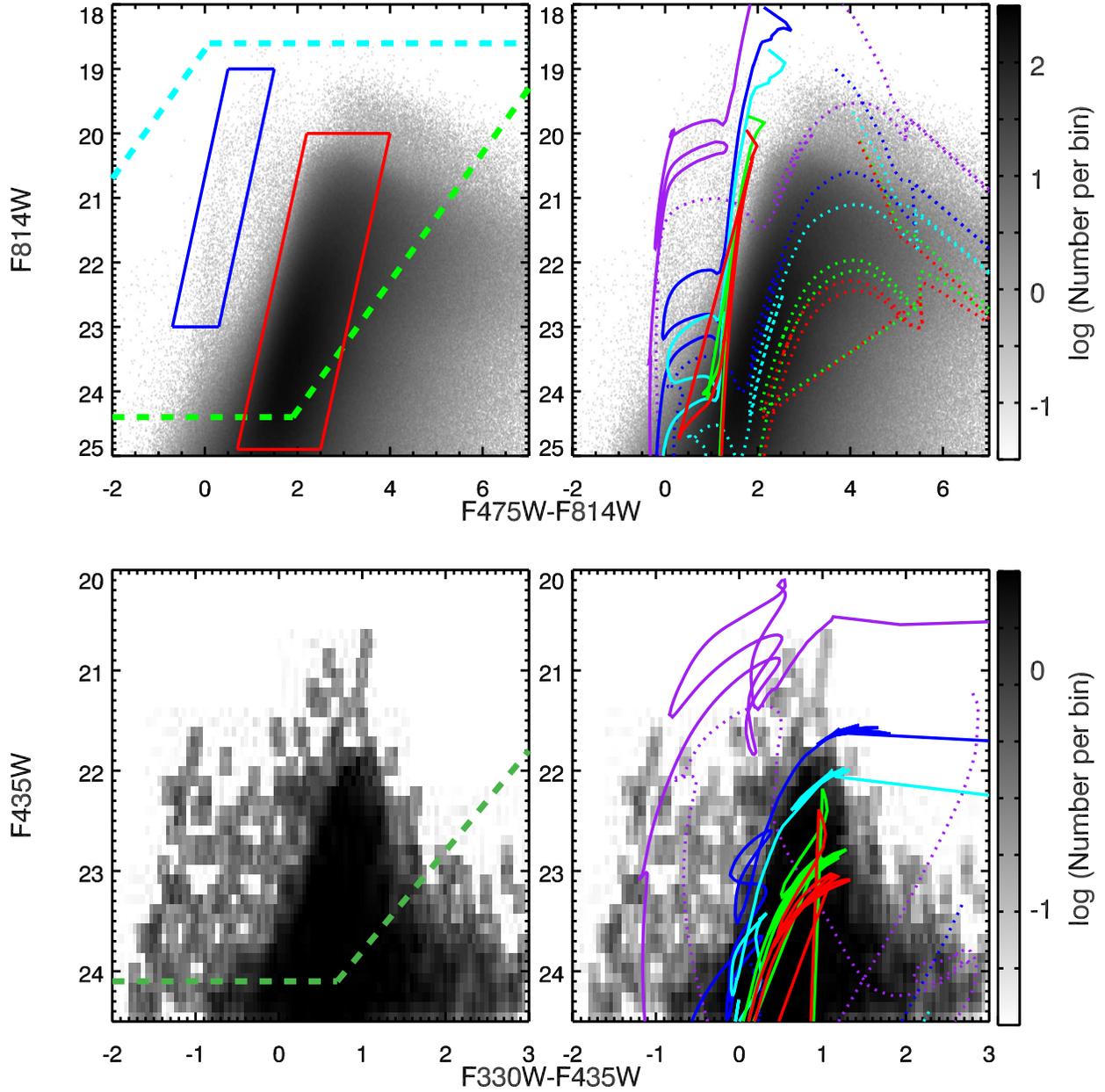,width=1.0\textwidth,angle=0}
       }
 \caption{The Hess diagram of the M31 bulge from the WFC (top) and HRC (bottom) observations. 
 On the top left panel, the cyan line marks the saturation limit of the WFC observation, 
 while the blue and red parallelograms embed blue and red stars, which 
 are discussed in \S\ref{ss:cmd}. 
 The green lines in the left column are the 50\% detection limit. 
 We overlay the isochrones from the Padova stellar evolutionary tracks in 
 the right column. The solid and dashed lines 
represent the stellar populations with 1/200 and 2.5 solar metallicities, respectively, 
at five different ages, 0.1 (purple), 0.5 (blue), 1 (cyan), 5 (green), 10 Gyr (red).}
\label{f:hess}
\end{figure*}

The Hess representations of the CMD of our ACS observations are shown in Fig.~\ref{f:hess} for 
both the WFC (top panels) and HRC (low panels) observations. 
%In the left column, the green dashed lines represent the 50\% detection limits. The 
%cyan lines on the top left panel 
%mark the saturation limit for the WFC observations. In the right column, 
%we overlaid the Padova isochrones of stellar population with 1/200 (solid) and 2.5 (dashed) 
%solar metallicity with different 
%ages~\citep[version: PARSEC v1.2S +COLIBRI PR16,][]{mar17,mar13,ros16}. 
We used 0.02 mag and 0.05 mag as the bin sizes of 
the color and magnitude terms in the CMD of the WFC observation, due to 
its large FoV, while 0.2 mag and 0.5 mag bin sizes are selected 
for the HRC observation. In the 
WFC observation, 
most stars concentrate in the regions enclosed in the red parallelogram, which 
are redder than most of the metal-poor isochrones and instead 
should be RGB stars of the metal-rich populations. On the 
other hand, the blue stars embedded by the blue parallelogram could be 
MS stars, %They are covered by both the metal-poor and metal-rich isochrones, 
%while for the same age, the former one is brighter than the latter one. 
%On the other hand, these stars could also be 
and/or post-AGB stars~\citep{ros12}. 
%and/or blue stragglers~\citep{mon11}. 
Like~\citet{mon11}, we called this region in the CMD `blue plume' (BP). 
%We will further discuss the origin of 
%these stars in \S\ref{ss:young}.  
In the HRC observation, most of the stars fall into the regions of the CMD 
close to or below the 50\% detection limit. On the other hand, the stars 
brighter than F435W=24 mag are divided in two branches:  
one with F330W-F435W$\sim$-1 and the other one with F330W-F435W$\sim$1. 
%could be explained by MS stars 
%(see \S\ref{ss:young}), 
%while the other one with F330W-F435W$\sim$0.3 seem to be brighter, even than 
%the isochrones of metal-poor population older than 500 Myr. (see \S\ref{ss:young}). 

\begin{figure*}[!thb]
  \centerline{
       \epsfig{figure=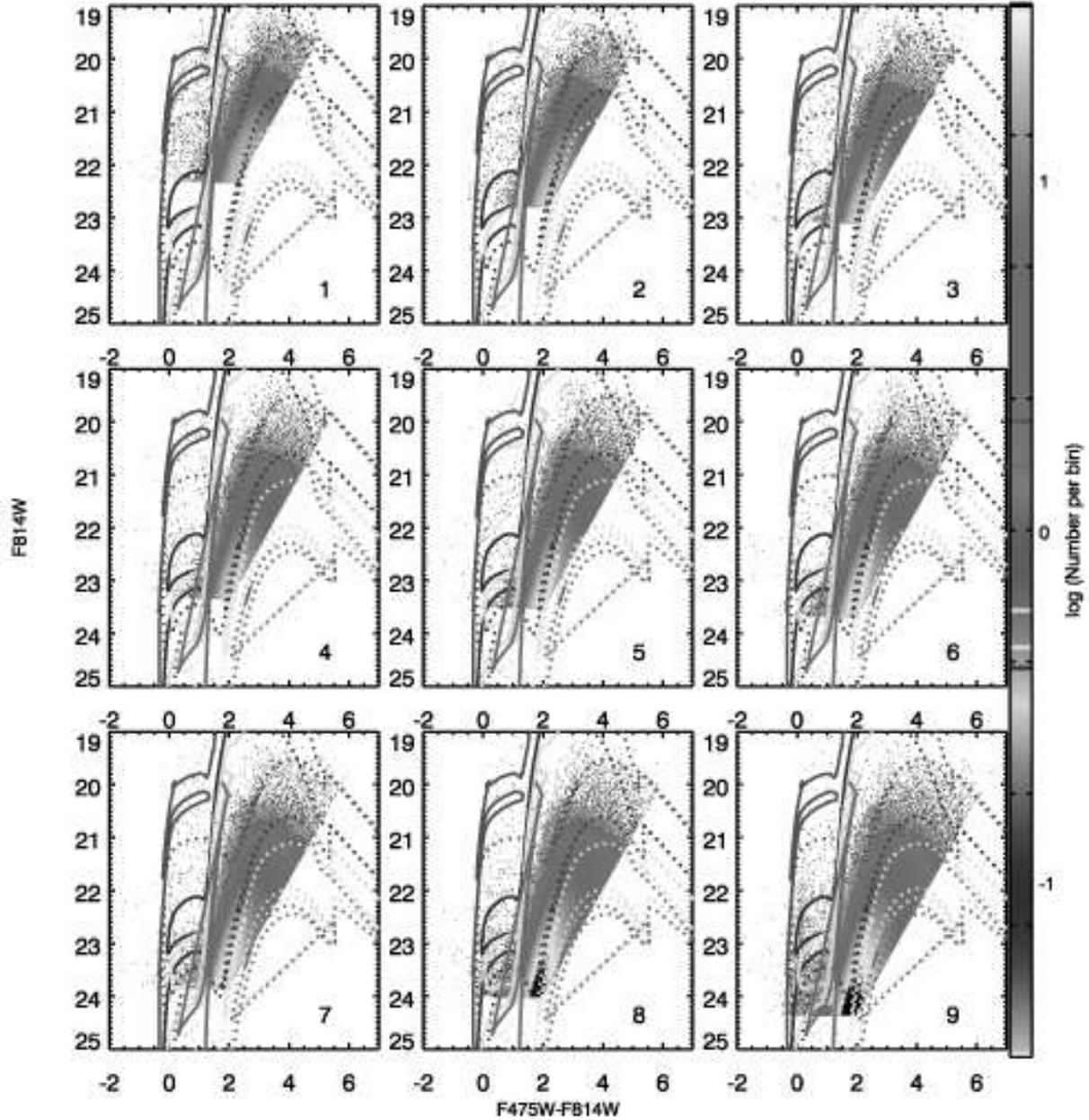,width=1.0\textwidth,angle=0}
       }
 \caption{The Hess diagram of the stars in the M31 bulge from the WFC observation at the nine annuli 
 defined in Table~\ref{t:annu}. Only the stars brighter than 50\% completeness limit at both bands are kept. 
  We also overlay the isochrones from the Padova stellar evolutionary tracks. The solid and dashed lines 
represent the stellar populations with 1/200 and 2.5 solar metallicities, respectively, 
at five different ages, 0.1 (purple), 0.5 (blue), 1 (cyan), 5 (green), 10 Gyr (red).}
\label{f:hess_pub}
\end{figure*}

We further present the evolution of the CMD along %the galactocentric 
%radius 
$R$ for the WFC observation. 
Fig.~\ref{f:hess_pub} shows the Hess representations of the CMD 
at the nine annuli 
defined in Table~\ref{t:annu}. In each annulus, only 
stars brighter than 50\% completeness limit are kept. %The same Padova 
%isochrones are overlaid. 
Due to the crowding, the recovered fraction in the innermost annulus is lowest. As a 
result, from the CMD, it is impossible to constrain the star formation rate (SFR) for the stellar 
population with super-solar metallicity with age older than $\sim$2 Gyr. 
On the other hand, the RGB stars with super-solar metallicity and 10 Gyr old 
are still brighter than 50\% completeness limit at both bands in annulus \#9. The BP 
stars with F475W-F814W$\sim0$ can be found in all the nine annuli, but mainly concentrate in 
annuli, \#8 and \#9 at F814W$\sim$23-24 mag. 

\begin{figure*}[!thb]
  \centerline{
       \epsfig{figure=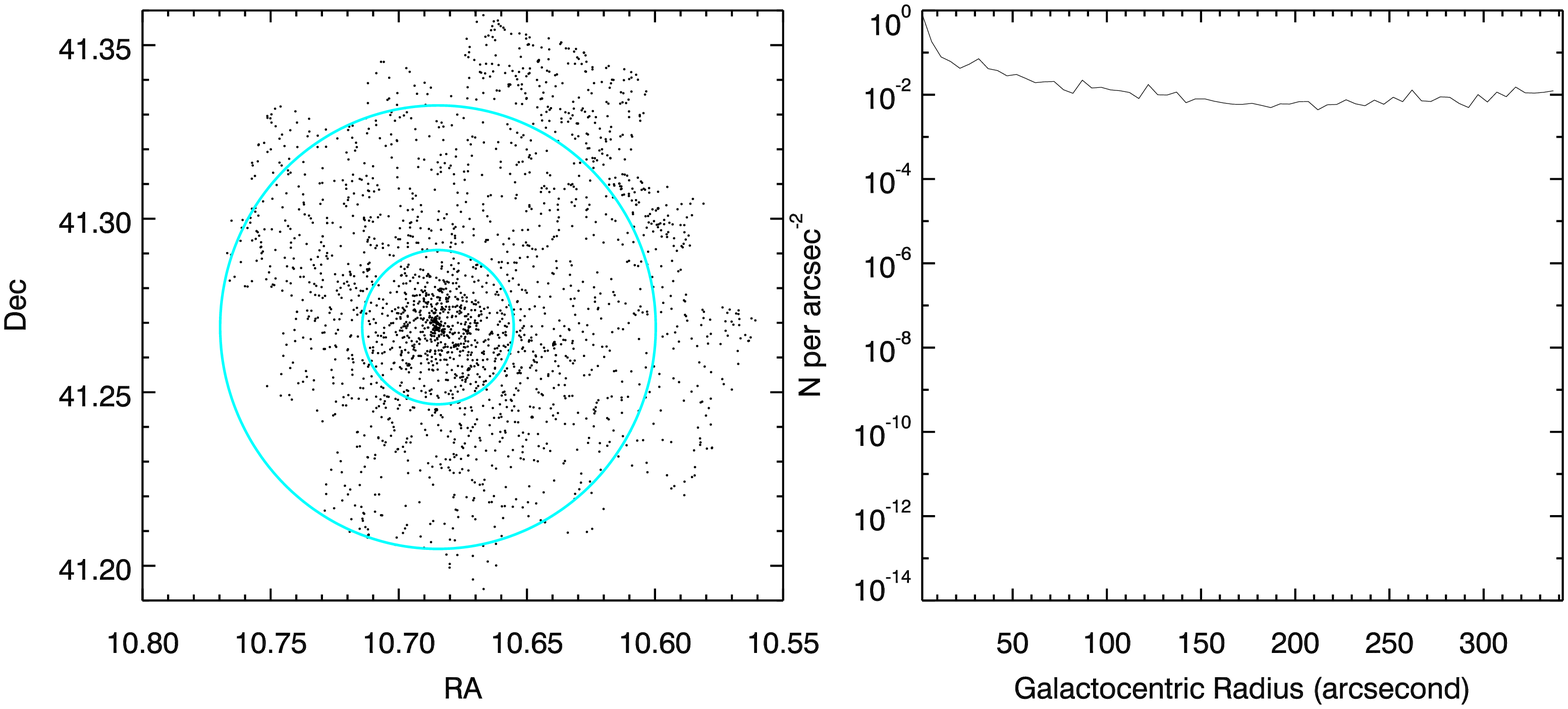,width=1.0\textwidth,angle=0}
       }
 \caption{The spatial distribution (left) and surface number density (right) 
 of sources within the blue parallelogram in 
 the top left panel of Fig.~\ref{f:hess}. 
 The two circles have galactocentric radii of 80\arcsec\ and 230\arcsec , respectively.
 For each galactocentric radius, the surface number density is the ratio of the number of stars in an annulus with 5\arcsec\ width to the effective sky area (arc second$^2$) within this annulus.}
\label{f:young_dis}
\end{figure*}

\begin{figure*}[!thb]
  \centerline{
       \epsfig{figure=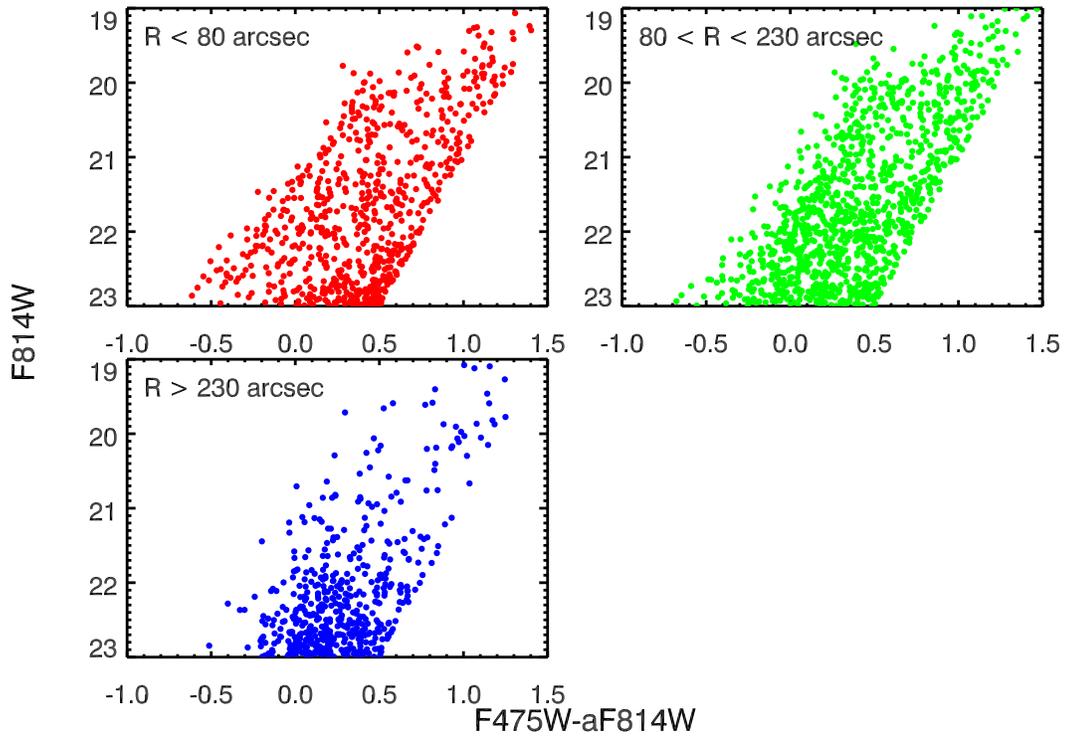,width=1.0\textwidth,angle=0}
       }
 \caption{The CMD of stars within the blue parallelogram in 
 the top left panel of Fig.~\ref{f:hess}, but at different galactocentric radii.}
\label{f:young_cmd}
\end{figure*}

%\subsubsection{Young Populations?}\label{sss:young}
In order to further study the nature of stars embedded within the blue parallelogram in 
the top left panel of Fig.~\ref{f:hess}, Fig.~\ref{f:young_dis} shows their spatial distribution. 
The majority of these stars concentrate in the M31 nucleus. Then, the surface number density 
decreases till $\sim$230\arcsec\ away from M31*, then increases again, caused by 
the enhancement of stars in the northwest of our FoV. %, i.e, where the 5 kpc star formation ring is. 
Fig.~\ref{f:young_cmd} shows the CMD at three different regions, divided by the two cyan 
circles with $R$ of 80\arcsec\ and 230\arcsec .
%Figs.~\ref{f:young_dis} and~\ref{f:young_cmd} show 
%their spatial distribution and their CMDs at three different regions, divided by the two cyan 
%circles with galactocentric radii of 80\arcsec\ and 230\arcsec . 
%For the stars beyond 230\arcsec , they show 
%an enhancement of stellar number density in the north 
%west of our field-of-view. 
In the lower left panel of 
Fig.~\ref{f:young_cmd}, the stars beyond 230\arcsec\ concentrate near  
$F475W$-$F814W\sim0.2$ and $F814W\sim$22.5. These 
stars should be MS stars in the 5 kpc star 
formation ring. 
Therefore, \#8 and \#9 annuli given in 
Table~\ref{t:annu} include not only the bulge 
stars, but significant disk contribution. On the other hand, the stars 
within the BP in the central 
230\arcsec\ distribute more uniformly in the CMD and do not seem 
to come from the young stellar population.%, but with a  
%potential enhancement at $F475W$-$F814W\sim0.5$ and 
%$F814W\sim$22-23, which can be seen in the CMD for the galactocentric 
%radius $<$ 80\arcsec\ (the top left panel of Fig.~\ref{f:young_cmd}). 

\begin{figure*}[!thb]
  \centerline{
       \epsfig{figure=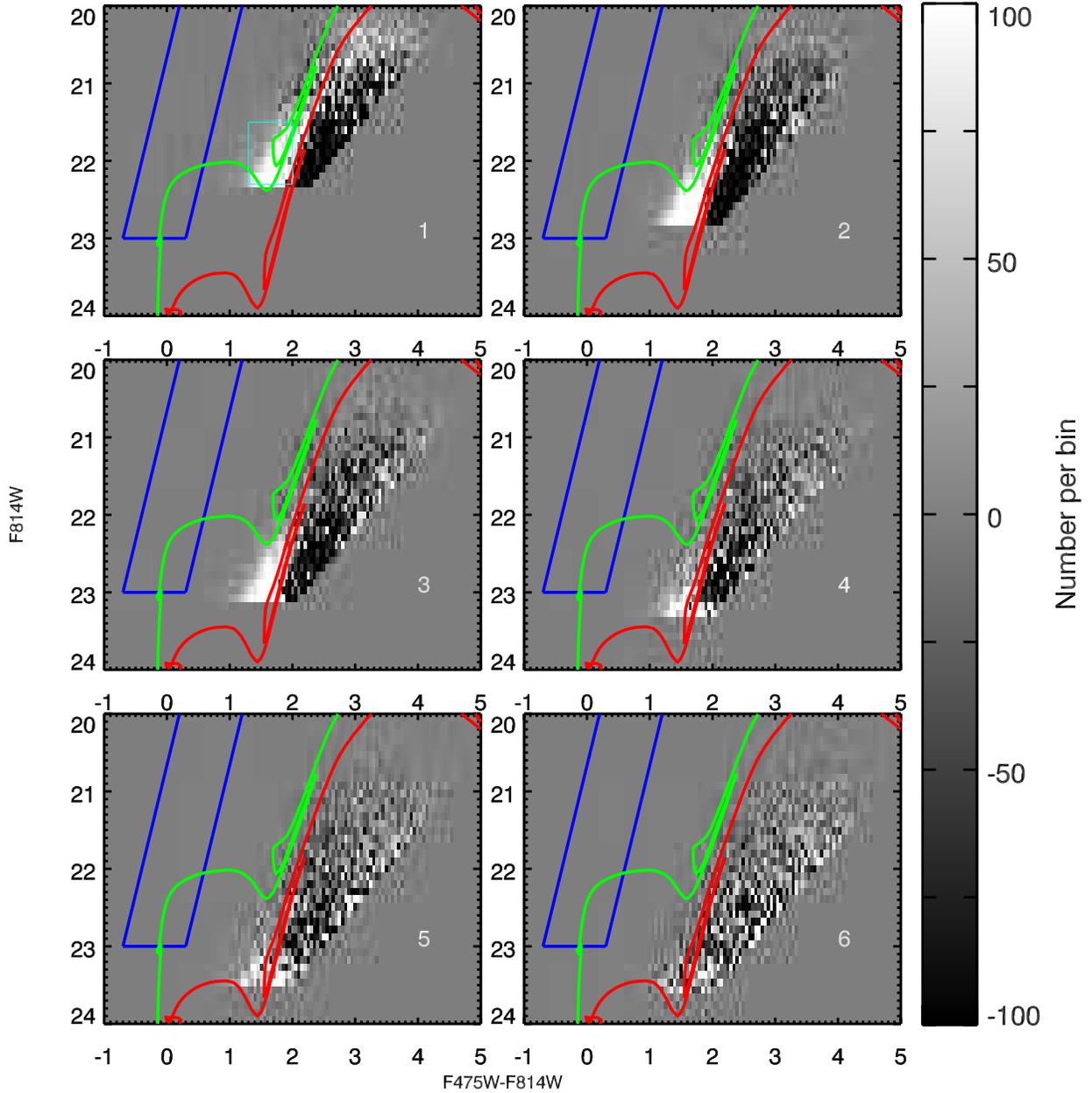,width=1.0\textwidth,angle=0}
       }
 \caption{The difference of the Hess diagrams between the six inner annuli and the \#7. 
 %The two outermost 
 %annuli are contaminated by the stars in the 5 kpc star formation rings. 
 The recovered fraction between the annuli is corrected. The CMD of the \#7 annulus have been scaled 
 to have the same total number of stars as the other annuli. The green and the red lines are the 200 Myr and 500 Myr old stellar populations with solar metallicity. The cyan box encompasses a region 
 in the CMD of the \#1 annulus with significantly extra stars relative to the \#7 annulus.}
\label{f:hess_pub_evo}
\end{figure*}

Fig.~\ref{f:hess_pub_evo} shows the differences of the CMDs 
between the six inner annuli and \#7 annulus to further demonstrate 
the evolution of the CMD in the M31 bulge. Unlike \#8 and \#9 annuli, 
these inner annuli should still be dominated by stars in the bulge. 
Before 
subtracting the CMD of \#7 annulus, we first corrected the difference of the 
recovered fraction among annuli and then scaled the CMD of \#7 annulus to 
the same total number of stars within the 50\% 
completeness limit of the individual inner annuli. Compared to the \#7 annulus, 
the innermost annulus has extra stars at F814W$\sim$22 mag and 
F475W-F814W$\sim$1.5, as marked by the cyan box. This region is redder than the stars in the BP, 
but can still be explained by the stellar population with solar metallicity and 200 Myr 
old. 
%Unfortunately, limited by the detection limit, the main sequence stars contributed 
%by this stellar population is still undetected. 
%the innermost annulus have more blue and bright stars, the number of which decreases 
%following the galactocentric radius. 
These stars become dimmer with the increase of $R$ from \#1 to \#4 annuli.

\subsection{Fitting the CMD}\label{ss:cmd_fit}
We used the source catalogs given in \S\ref{ss:decon}
and the artificial star tests performed in \S\ref{ss:artificial}
to constrain the stellar populations in the M31 bulge. Specifically, we 
utilized the CMD fitting method developed by~\citet{ols06} 
to compare the observed data with those predicted by the stellar 
evolutionary isochrones with different ages and metallicities. This method first uses 
the stellar evolutionary tracks to generate CMDs, giving a set of parameters (e.g. 
age, metallicity, distance, reddening, %binary properties 
and initial mass function). The synthetic 
CMD is convolved with the observed photometric uncertainty to simulate the real 
observation. The likelihoods of the observed CMD as a representation of a set of starburst 
stellar populations are calculated and are used to determine 
the best solution. 
%. Specific technique is used to searching the best solution
%in the likelihood space. 

The Padova stellar isochrones were utilized to 
simulate the intrinsic CMD. 
We produced a series of instantaneous star burst with different ages and metallicities. 
The ages were from the set {125 Myr, 375 Myr, 605 Myr, 855 Myr, 1.2 Gyr, 1.7 Gyr, 
2.4 Gyr, 3.4 Gyr, 4.8 Gyr, 6.8 Gyr, 9.6 Gyr, 13.5 Gyr}. The metallicity were from the 
set Z={0.003, 0.008, 0.019, 0.03, 0.05}. These choices made sure that our grids were large enough to 
cover  big age range and were small enough to discover the different features in the CMD. 
%For the young 
%ages ($<$1 Gyr), the high-mass stars evolve very quickly in the CMD. Small 
%age steps are used to recover the different stellar types. Instead, after $<$1 Gyr, the high-mass 
%stars have died as supernova, while the low-mass stars are still in the main-sequences. These stars 
%evolve much slowly and a age step, $\sim$ 2 Gyr, is enough to capture the evolution of the 
%CMD. 

The molecular clouds in the M31 bulge could 
complicate our investigation; we did not 
know the relative radial locations of individual stars, respective to the 
molecular clouds, so that we did not know how much foreground 
extinction which these stars suffered from. Fortunately, the M31 bulge is lack of 
molecular clouds~\citep{li09,mel11,don16}. In order to test the effects 
of the molecular clouds on our CMD fitting results, we 
performed the CMD fitting on the catalogs with or without the regions of high 
dust surface density. Specifically, 
we used the \herschel\ dust map~\citep{gro12} to identify the regions 
with large dust mass, i.e. the regions with 
$\sum_{dust}>$0.1 M pc$^2$. By using the dust mass to visual extinction ratio 
from~\citet{dra13} and the relative extinction from~\citet{don14}, this corresponded  
to A$_{F814W}>$0.23 and A$_{F475W}$-A$_{F814W}>$0.23. 
Only 24\% of our FoV satisfied this requirement. 
%Although the 
%line-of-sight with large dust mass could 
%have a molecular clouds far behind the M31 bulge and 
%would not affect most of the detected stars~\citep{mel11}, 
%removing these regions help us to reduce the 
%uncertainty of dust in our analysis. 
 
For the other parameters: 1) We adopted the distance modulus, 
24.46 (780 kpc,~\citealt{mcc05,gri14});  
2) We chose the~\citet{kro01} IMF. 
The exact choice of the IMF was not highly 
relevant because the different standard IMFs suggested in 
the literature differ mainly at the low stellar mass range, which lied  
far below our detection limit. As a result, 
alternative IMF could only alter the total mass of the system, but not the relative 
SFH;  
3) The foreground Galactic extinction (A$_{F330W}$=0.282 mag, A$_{F435W}$=0.226 mag, A$_{F475W}$=0.203 mag and 
A$_{F814W}$=0.095 mag)    
was considered in the CMD fitting, while no extra local extinction was included. 

We performed the CMD fitting for the WFC observation at the 
nine annuli defined in \S\ref{ss:artificial} and the HRC observation. 
We divided our fitting 
processes into several steps to see the contribution of stellar populations with different ages in 
the observed CMD. Considering that the M31 bulge is known to be dominated by old stars, we first 
fit the observed CMD with models older than 5 Gyr. Then, we added the stellar populations older than 
1 Gyr, but younger than 5 Gyr. At the end, we used all the stellar populations. 

The uncertainties of the CMD fitting include two parts: statistic and 
systematic errors. 1) The statistic uncertainty comes from the Poisson 
fluctuation of number of stars in each bin of the CMD. To determine 
this uncertainty, we add random errors into individual bins of 
the CMDs of the nine annuli and fit 
the new CMDs. For each annuli, this procedure has been repeated for 1000 times. The 
standard deviation of the output parameters are used to be the 
statistic uncertainty of the CMD fitting; 2) The systematic uncertainty is 
introduced by the distance, foreground extinction and the bin effect. 
%The technique given in~\citet{mon12} is used to calculate the uncertainties of the CMD fitting, 
%which is dominated by the systematic uncertainty introduced by the distance, local extinction 
%and the bin effect. 
The uncertainty of the distance is 0.1 mag~\citep[-0.1, 0 and 0.1 mag,][]{gri14}. 
For the local extinction, 
we add an extra extinction of $A_{V}$=0, $\pm$0.06 and  
$\pm$0.12 into the fitting and assuming 
the extinction curve %in the M31 bulge 
given by~\citet{don14}. 
For the bin effect, we shift in color and magnitude with one third of the bin size 
(-1/3, 0 and 1/3 of the bin size). 
Then, we rerun the CMD fitting for these, 3 (distance) $\times$ 5 (local extinction) 
$\times$ 9 (bin shift) 
= 135, new dataset and derived the SFR at different ages and metallicities. The 
68\% percentiles of the values of these 135 new dataset are used to be 
the systematic uncertainty of the SFR derived above. 

\begin{figure*}[!thb]
  \centerline{
       \epsfig{figure=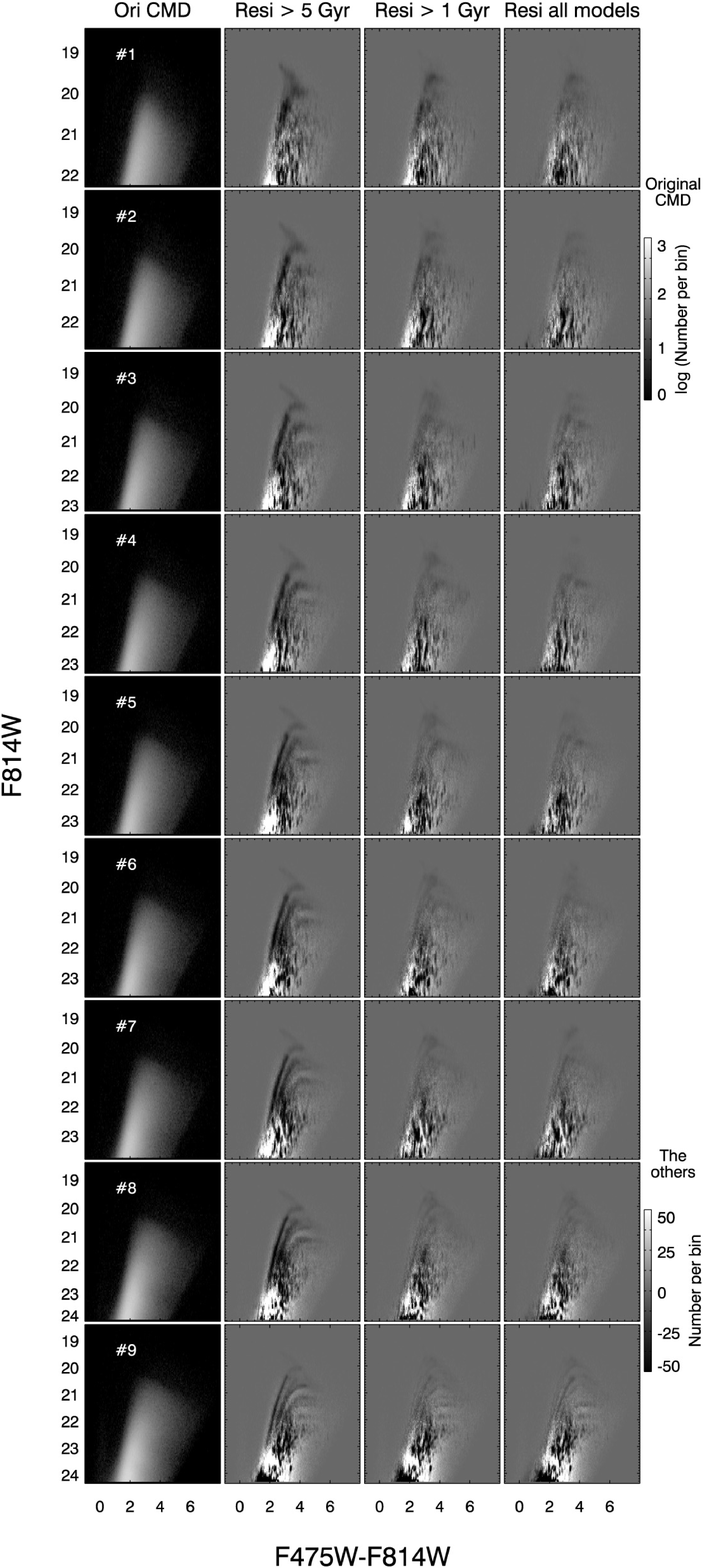,width=0.6\textwidth,angle=0}
       }
 \caption{The CMD fitting result for the WFC observation. The first column is the 
 observed CMD. The second, third and fourth columns are the 
 fitting residual using the models with $>$ 5 Gyr only, $>$ 1 Gyr only and 
 all the ages.}
\label{f:fit_result}
\end{figure*}

\begin{figure*}[!thb]
  \centerline{
       \epsfig{figure=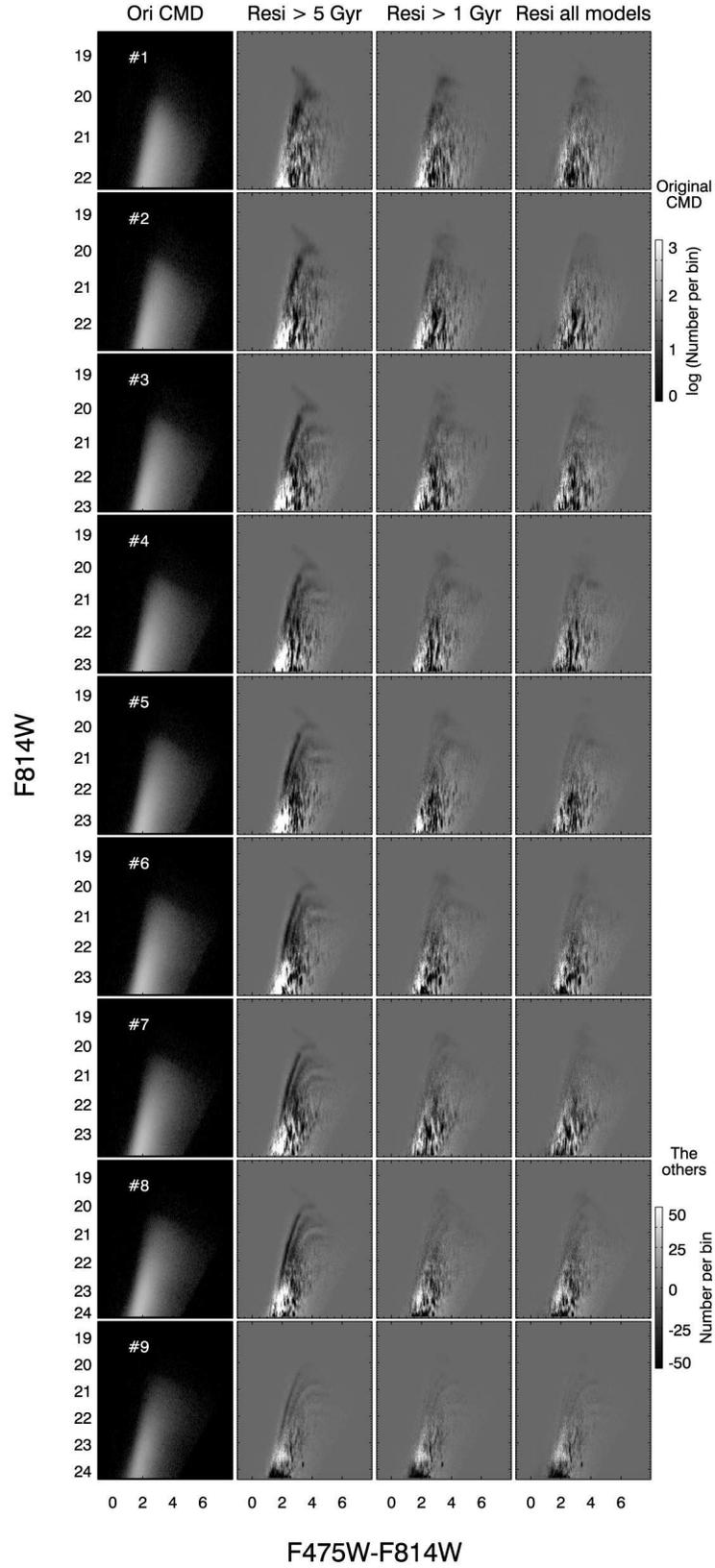,width=0.6\textwidth,angle=0}
       }
 \caption{Similar to Fig.~\ref{f:fit_result}, but for the regions 
 without high extinction (see \S\ref{ss:cmd_fit}).}
\label{f:fit_result_nex}
\end{figure*}

Figs.~\ref{f:fit_result} and~\ref{f:fit_result_nex} show the final fitting results 
of the three steps for the WFC observation at the nine annuli with or without the 
high extinction regions. Overall, these two figures are very similar; That is because 
the high extinction regions only occupy a small fraction of our FoV. 
In the second column for the fitting with stellar populations older than 5 Gyr, significant 
residuals exist at the regions with $F475W$-$F814W$=2 and $F814W$$\sim$23. They 
disappear with the inclusion of younger stellar populations. Compared the third 
and fourth columns, we see the potential existence of young stars with age younger than 1 Gyr, 
especially for the \# 1 and \#2 annuli. % between \#1 to \#2. 

\begin{figure*}[!thb]
  \centerline{
       \epsfig{figure=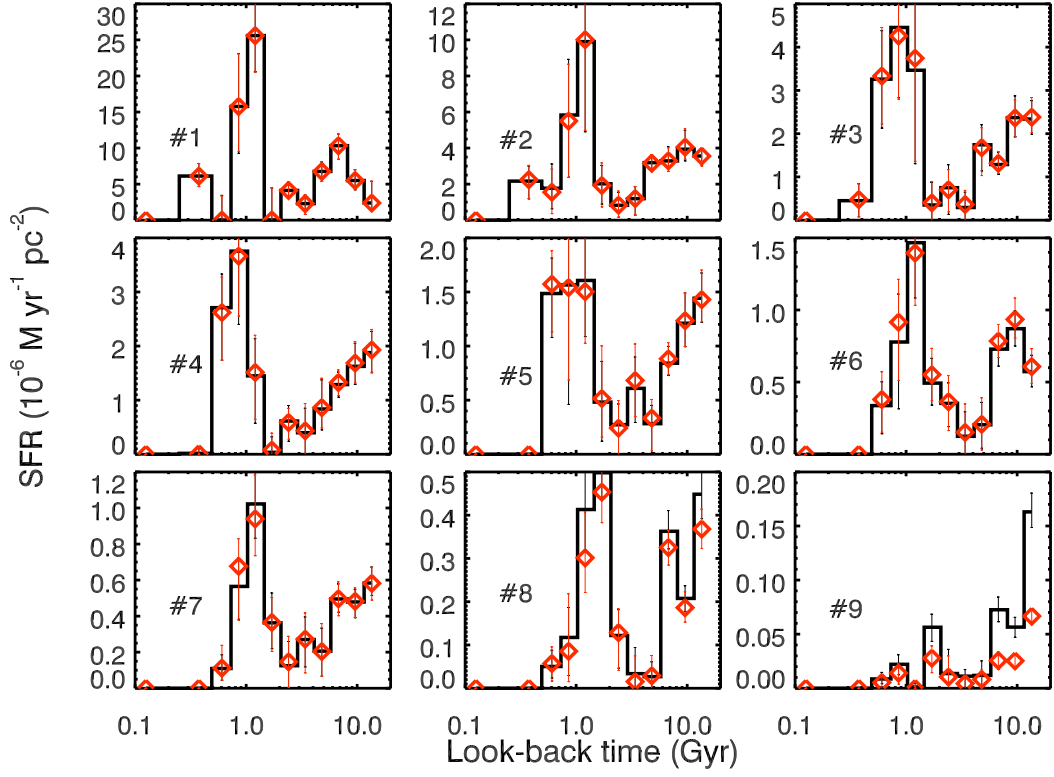,width=1.0\textwidth,angle=0}
       }
 \caption{The star formation history for the nine annuli derived 
 from the CMD fitting with (solid lines) or without high-extinction regions 
 (red diamonds, see \S\ref{ss:cmd_fit}).}
\label{f:sfh_evo}
\end{figure*}

\begin{figure*}[!thb]
  \centerline{
       \epsfig{figure=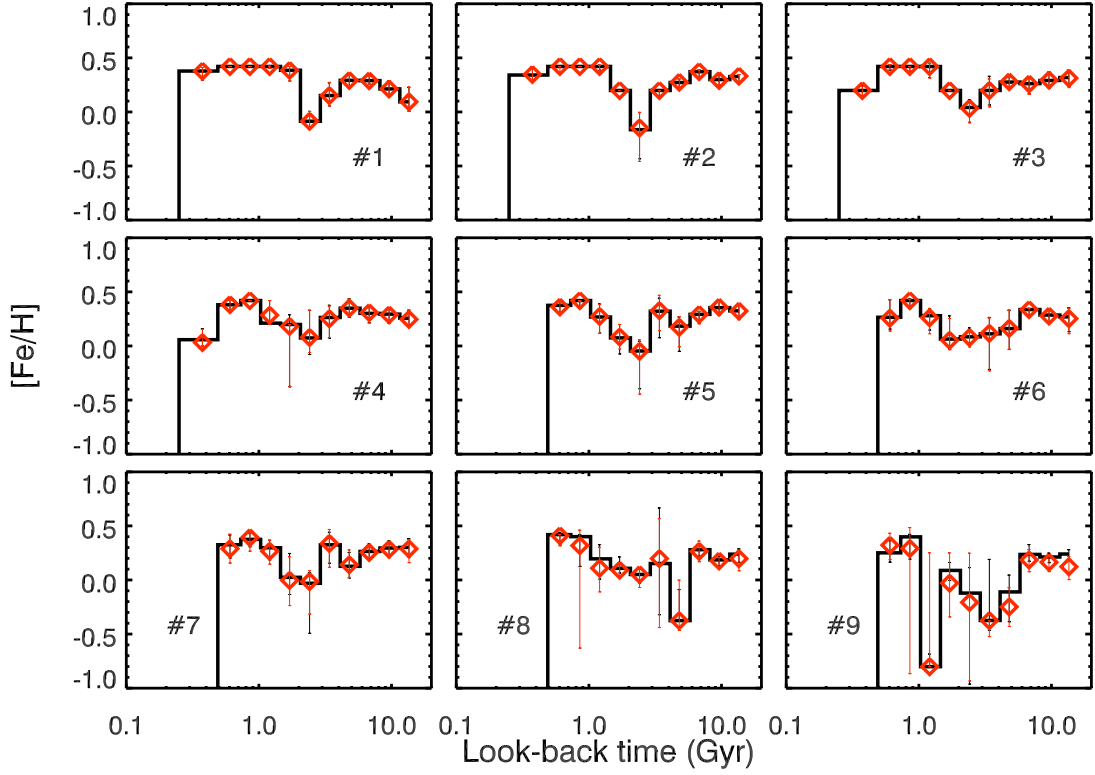,width=1.0\textwidth,angle=0}
       }
 \caption{The metallicity evolution for the nine annuli derived 
 from the CMD fitting with (solid lines) or without high-extinction regions 
 (red diamonds, see \S\ref{ss:cmd_fit}).}
\label{f:me_evo}
\end{figure*}

\begin{figure*}[!thb]
  \centerline{
       \epsfig{figure=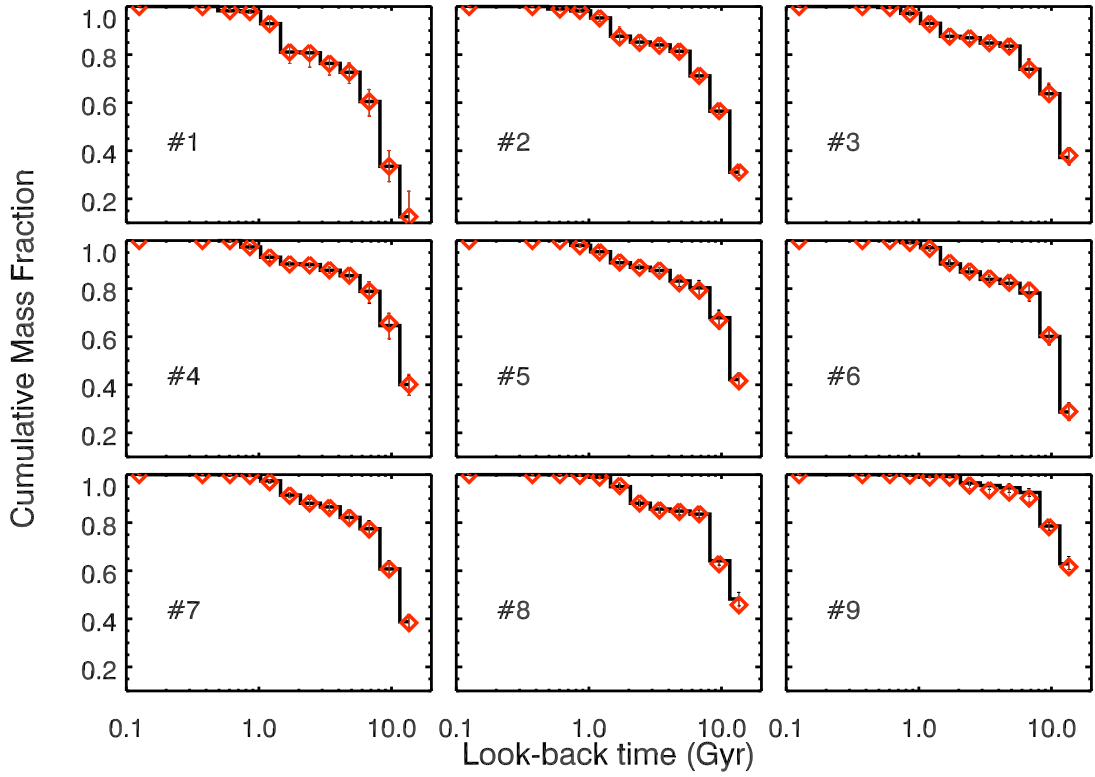,width=1.0\textwidth,angle=0}
       }
 \caption{The cumulative mass-weighted age distribution for the nine annuli derived 
 from the CMD fitting with (solid lines) or without high-extinction regions 
 (red diamonds, see \S\ref{ss:cmd_fit}).}
\label{f:cum_evo}
\end{figure*}

\begin{figure*}[!thb]
  \centerline{
       \epsfig{figure=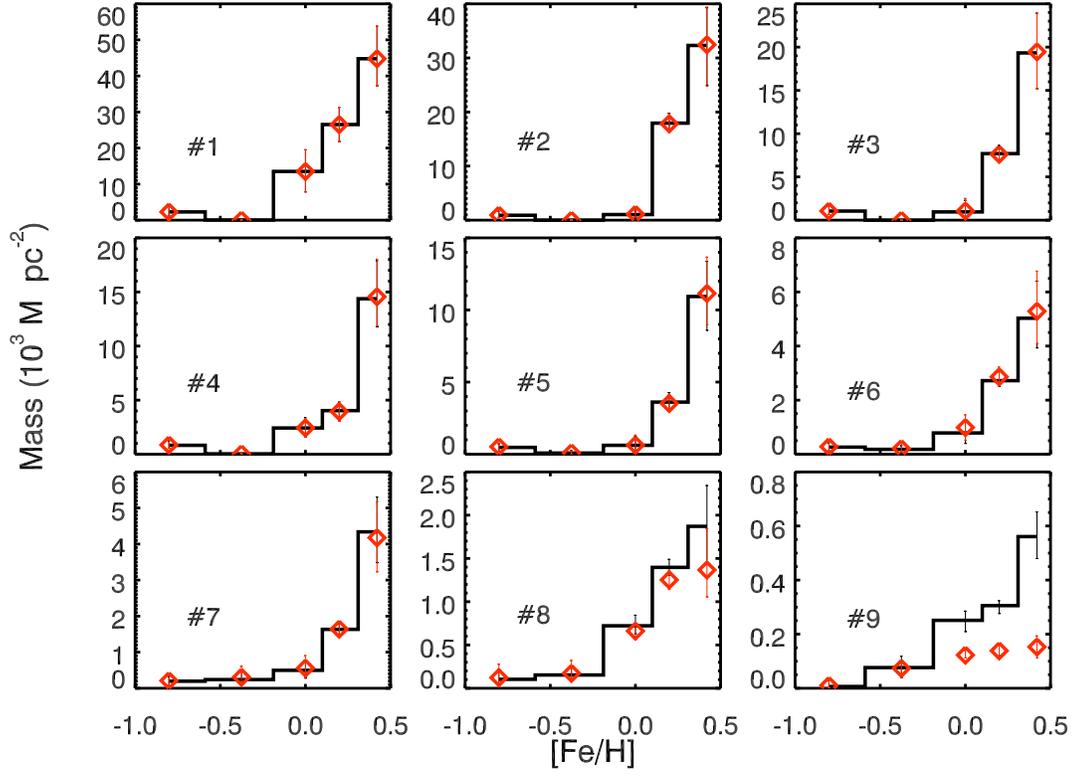,width=1.0\textwidth,angle=0}
       }
 \caption{The mass as a function of metallicity for the nine annuli derived 
 from the CMD fitting with (solid lines) or without high-extinction regions 
 (red diamonds, see \S\ref{ss:cmd_fit}).}
\label{f:me_dis}
\end{figure*}

\begin{figure*}[!thb]
  \centerline{
       \epsfig{figure=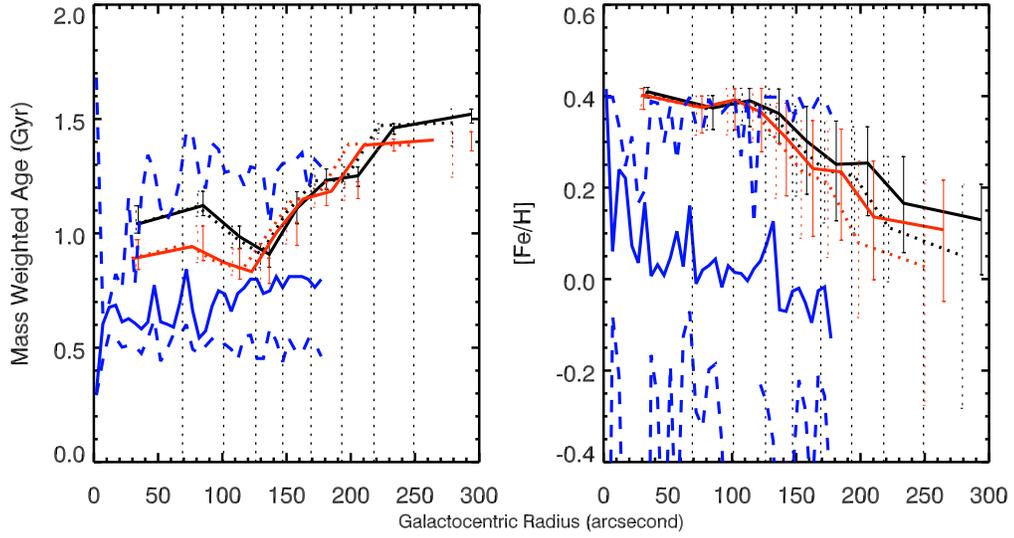,width=1.0\textwidth,angle=0}
       }
 \caption{The evolution of the mass-weighted (black lines) 
 and luminosity-weighted (red lines) age (left) and metallicity (right) for the 
 stellar populations younger than 2 Gyr at the nine annuli. The black/red 
 solid and dashed lines are for the CMD fitting with and without 
 the high-extinction regions (see \S\ref{ss:cmd_fit}). The error bars include 
 both statistic and systematic errors in the CMD fitting described in 
 \S\ref{ss:cmd_fit}. The blue solid and 
 dashed lines are the age/metallicity and their uncertainties of 
 the intermediate-age stellar population given in~\citet{don15}. The eight 
 black vertical dotted lines are the boundaries of the nine annuli 
 defined in Table~\ref{t:annu}.}
\label{f:age_me_r_evo}
\end{figure*}

Figs.~\ref{f:sfh_evo},~\ref{f:me_evo},~\ref{f:cum_evo} and~\ref{f:me_dis} show the results of the 
CMD fitting of the nine annuli: SFR as a function of look-back time; the 
metallicity as a function of look-back time; the cumulative mass-weighted age distribution and 
the mass as a function of metallicity. In all the annuli, there is a star formation activitiy 
around 1 Gyr ago, although it slightly shifts to the older age: This evolution is clearly shown 
in Fig.~\ref{f:age_me_r_evo}, the mass-weighted age changes from $\sim$1 Gyr in the 
innermost annulus to $\sim$1.5 Gyr at the outermost annulus. The relative contribution from 
the old stellar population increases toward the outer regions, which is due to the detection limit. 
In the inner region, we can see the recent star formation activity ($<$ 500 Myr) happened in 
the two innermost annuli, but are not significantly beyond that. Overall, the metallicity in the M31 
bulge is high for all the stars. 

\begin{figure*}[!thb]
  \centerline{
       \epsfig{figure=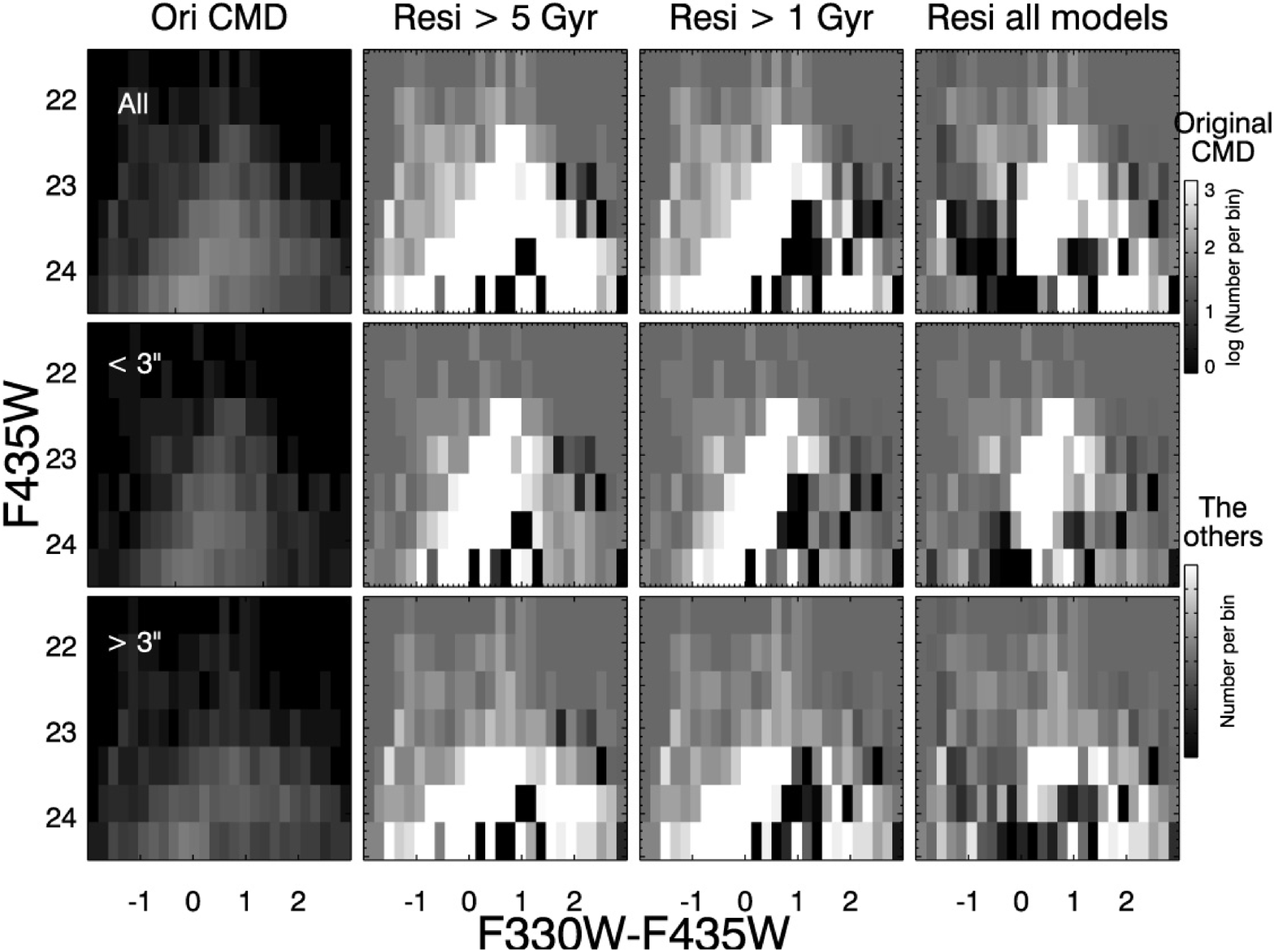,width=1.0\textwidth,angle=0}
       }
 \caption{The CMD fitting result for the HRC observation. The first column is the 
 observed CMD. The second, third and fourth columns are the 
 fitting residual using the models with $>$ 5 Gyr only, $>$ 1 Gyr only and 
 all the ages.}
\label{f:fit_result_hrc}
\end{figure*}

\begin{figure*}[!thb]
  \centerline{
       \epsfig{figure=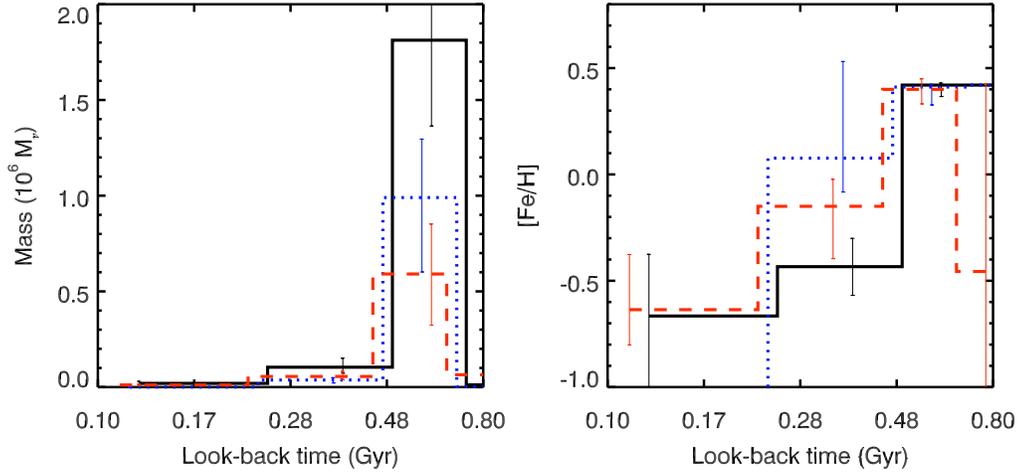,width=1.0\textwidth,angle=0}
       }
 \caption{Left: total stellar mass as a function of look-up time up to 0.8 Gyr; 
 Right: age-metallicity distribution. The black, blue and red lines are 
 for all the stars, the stars inside and outside $R$=3\arcsec\ of the HRC observation.}
\label{f:hrc_cmd_output}
\end{figure*}

The top panel of Fig.~\ref{f:fit_result_hrc} shows the result of the same CMD fitting 
for the HRC observation. However, we see that even in the last column using models 
with all the ages and metallicities, the model can not explain the stars with 
F330W-F435W$\sim$1 and F435W$<$23.5. In order to check whether these stars are 
blending stars, we divide the catalog into two regions using $R$=3\arcsec\ (see Fig.~\ref{f:hrc_f330w}). 
The middle 
and low panels of Fig~\ref{f:fit_result_hrc} shows their CMD fitting results, respectively. 
However, we can see that 
there is still many bright and red stars in the CMD outside the 3\arcsec , which can not 
be explained by the blending. Fig.~\ref{f:hrc_cmd_output} gives the produced stellar mass 
and the metallicity as a function of look-back time during the last 800 Myr for 
the whole HRC observation (black solid lines), the inner 3\arcsec\ (blue dotted lines) and beyond 
(red dashed lines). Because the shallowness of the HRC observation, we do not trust the SFR 
for the stellar population older than 1 Gyr; for example, in Fig.~\ref{f:hess}, although 
for the extremely metal-poor metallicity (1/200 Z$_{\odot}$), we can detect the RGB stars for 
the 10 Gyr stellar population (red solid line), we cannot detect any stars from the 1 Gyr old 
stellar population (cyan dotted line). Fig.~\ref{f:hrc_cmd_output} tells us in the central 14\arcsec\ (the 
FoV of HRC camera), the central 3\arcsec , and between 3\arcsec\ and 14\arcsec , the star formation activity produced 1.0/0.4/0.6$\times10^5$ and 1.8/1.0/0.6  $\times10^6$ M$_{\odot}$ at around 0.375 and 0.6 Gyr ago. %These stars have 
The stars that formed at 0.375 Gyr ago have 
around or sub-solar metallicity ([Fe/H]$<$0). 

%\textbf{What is the TRGB magnitudes of M31 and whether they are TPAGB in M31 to identify young 
%stellar population (~ 1Gyr)}
%\textbf{According to Fig. 3 of Rosenfield et al. 2016, for TP-AGB, F160W $<$ -6, F814W-F160W$\sim$1.5, 
%i.e. F814W$\sim$-4.5, considering the distance moduli of M31, 24.46, i.e. F814W=19.96. The TP-AGB 
%stars exists in all field-of-view.}

\subsection{Blending?}\label{ss:blending}
We have analyzed a very crowded field: the M31 bulge. As a result, our source 
catalogs could suffer  
from the blending effect.  
%Two dim stars could blend together to 
%become a brighter star. 
%As mentioned in the Introduction, the blending effect 
%seriously hamper 
%the attempt to accurately study the real star formation 
%history in previous work. 
%Underestimating 
%or ignoring this effect could cause us to obtain wrong results. 
Compared to the HRC 
observation,
the WFC observation is more affected by the crowding, 
 because 
of the worse resolution and larger surface brightness. % contributed by 
%low-mass stars at the optical bands. %, compared to 
%the near-UV bands. 
Furthermore, the blending is more significant 
at the F814W band than the F475W band, which causes stars artificially brighter and 
slightly redder (see also \S\ref{ss:artificial_cluster}). 
%the F475W band, and even worse at the 
%F814W band, which will cause stars artificial brighter and 
%slightly redder (see also \S\ref{ss:artificial_cluster}). 
% Meanwhile, it is also hard to 
%let the CMD fitting in \S\ref{ss:cmd_fit} to take 
%this effect into account. This effect is more. 
Therefore the blending could potentially explain the differential CMDs 
between the inner six annuli and the \#7 annulus shown in Fig.~\ref{f:hess_pub_evo}, 
instead of the intrinsic evolution of stellar populations along $R$. 

%In the next two subsections, 
%we will use two different methods to examine the blending effect on our results. 

   %Blending effect has serious impact on the regions in 
%the CMD occupied by the AGB stars, because these stars are rare and have  
%a significant weight on determining the star formation rate 
%around the intermediate-age stellar populations. 

\begin{figure*}[!thb]
  \centerline{
       \epsfig{figure=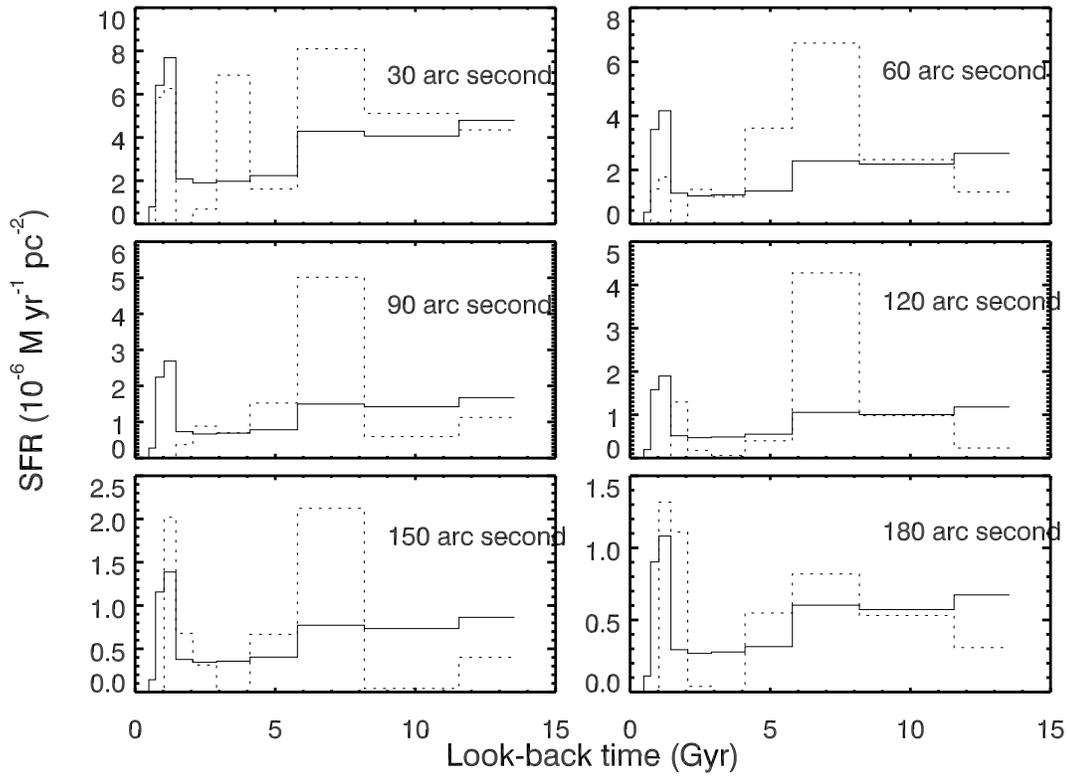,width=1.0\textwidth,angle=0}
       }
 \caption{The CMD fitting outputs from the artificial observation with the surface 
 densities equal to those at different radii. The solid line is the input SFH, 
 while the dotted line is the output SFH.}. 
  \label{f:blending_simul}
\end{figure*}

%\subsubsection{Complete artificial observations}
We used the Padova stellar isochrones and the PSF determined in 
\S\ref{ss:decon} to simulate WFC observation to examine the blending effect.  
We scaled the SFH of the \#7 annulus determined in \S\ref{ss:cmd_fit} 
to let the surface densities of artificial observation 
match those at $R$ from 30\arcsec\ to 180\arcsec\ with a step size of 
30\arcsec . Then, for each $R$, we produced an artificial source
catalog, multiplied 
the intensities of individual sources with the PSF and added them into a blank image. 
Poisson noise has been added into this image. 
After that, from this artificial image, we produced the deconvolution catalog and performed 
the CMD fitting. Then, we compared the output SFHs with the input one at different 
radii, which are shown in Fig.~\ref{f:blending_simul}. 

From this plot, we can see that even in the artificial observation with the highest 
surface densities, i.e. $R$=30 arc second, 
the output CMD fitting does not overestimate the SFH at $<$ 2 Gyr: 
especially we do not see any significant star formation activities 
around several hundred Myr, compared to the input CMD. Therefore, the 
blending effect can not explain the $<$ 1 Gyr old stellar population in 
the central four annuli mentioned in \S\ref{ss:cmd_fit}. On the other hand, the output CMD fitting 
has higher SFR around 5-6 Gyr. The stars at this age range are 
near the detection limit. Therefore, the blending can cause dim stars with 
older ages ($>$ 5 Gyr) brighter and let the CMD fitting 
obtain higher SFR at the 5 Gyr year old range.

%\subsection{Metallicity of the M31 bulge}\label{ss:rgb}

\section{Discussion}\label{s:discussion}
We first %discuss the blending effect on our 
%source catalogs and the CMD fitting 
%in \S\ref{ss:blending}. 
%Then, we 
%classify whether the stars fall onto  
%the BP of the WFC observation 
%(F814W$\sim$22 mag and 
%F475W-F814W$\sim$1.5) 
%and the blue stars with F330W-F435W$<$-0.3 in the 
%HRC observation are young MS stars or 
discuss whether $\sim$200 Myr old stars exist in the 
M31 bulge or not in \S\ref{ss:blueplume}. 
Then, we analyze the stellar populations and their spatial 
evolutions in the M31 bulge in \S\ref{ss:population}. %After that, 
%we explain how the M31 bulge formed in \S\ref{ss:formation}. 
In \S\ref{ss:formation}, 
we present the star formation history of the M31 bulge derived from our data.
At the end, we compare our results with previous ones derived from 
the SED fitting of photometric observation in \S\ref{ss:com_sed} to examine the reliability 
of SED fitting on the study of stellar populations in remote galaxies. 

%the stellar population in 
%the blue plume and the potential young stars  . At the end, we will put our results 
%into the discussion of the formation of the M31 bulge in \S\ref{ss:formation}. 

%\subsection{Young stellar population in the M31 bulge?}\label{ss:young}
%In \S\ref{ss:cmd}, we identify two unique regions in the our CMD, the blue 
%plume and the cyan box in the top left panel of Fig.~\ref{f:hess_pub_evo}. 
%We will separately discuss the origin of stars fall into this two subsection below. 

\subsection{$\sim$200 Myr old stars in the M31 bulge?}\label{ss:blueplume}
Fig.~\ref{f:hess}  shows that 1) there is a handful of stars that fall into the BP region 
of the CMD derived from the WFC catalog; 2) a number of stars with 
F330W-F435W$<$-0.3 in the HRC catalog. These stars 
could potentially be MS stars. However, on the other hand, 
evolved low-mass stars in the P-AGB, PE-AGB, AGB-manque  
or even the Planetary Nebulae (PNe) stage, 
could also be very blue, because 
%the materials 
%in their surface has been removed by the stellar wind. 
they have lost the majority or all of their envelopes in the form of 
strong stellar winds. Therefore, we perform two steps below 
to further understand the origin of these blue stars. 

\begin{figure*}[!thb]
  \centerline{
       \epsfig{figure=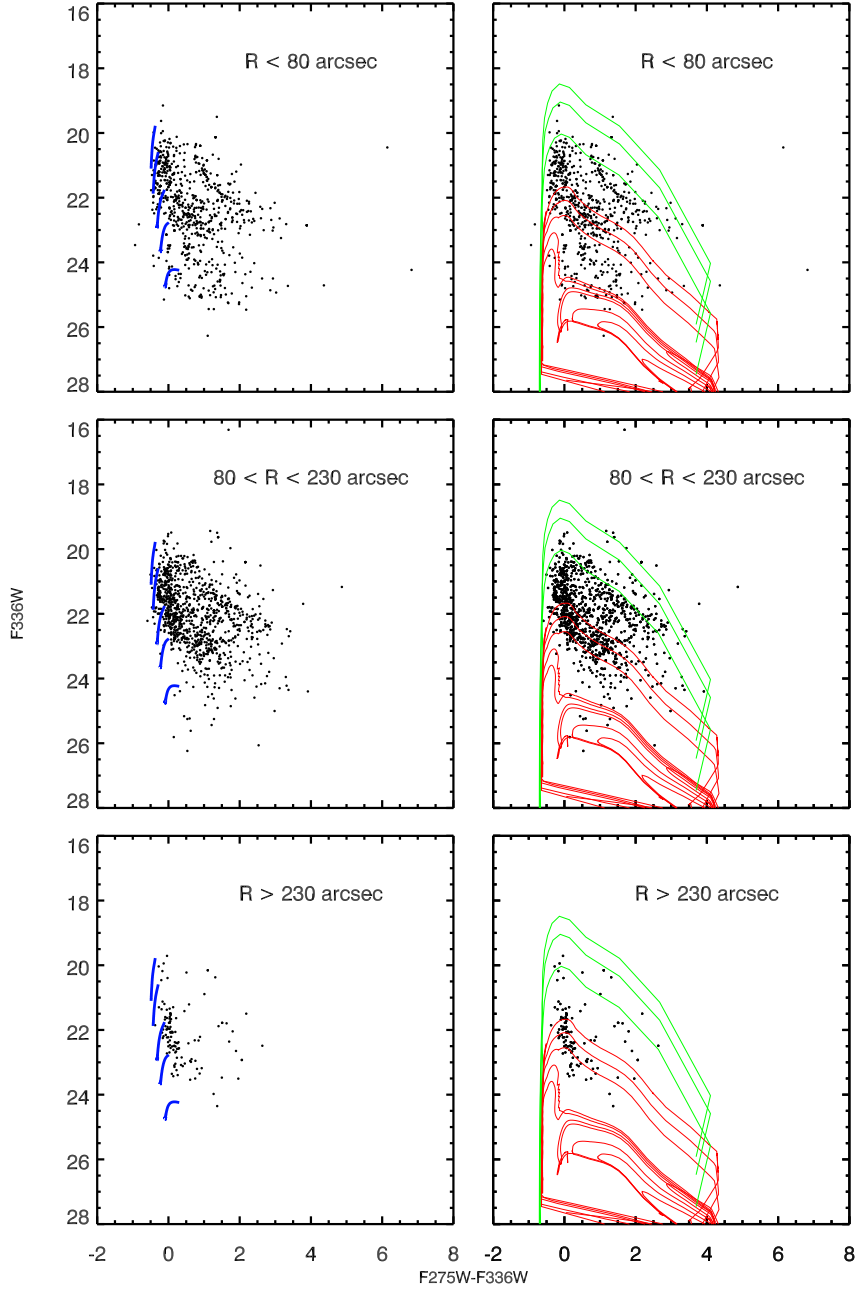,width=0.8\textwidth,angle=0}
       }
 \caption{The CMD (F275W-F336W vs F336W) of stars fall in the blue 
 parallelogram %(left: blue parallelogram and right: red parallelogram) 
 of Fig.~\ref{f:hess} at three different regions (from top to bottom panels) defined 
 in Fig.~\ref{f:young_dis}. In the left and right columns, we compared the CMD 
 with the MS tracks for young stars with 9, 7, 5, 4 and 3 solar mass (left) and 
 the tracks of the evolved low-mass stars at the P-AGB (green lines, from 
 top to bottom, 0.9, 0.75 and 0.6 $M_{\odot}$) and HP-HB (red lines, from 
 top to bottom, 0.58, 0.56, 0.54, 0.5 and 0.48 $M_{\odot}$.) 
 phases (right), as 
 described in~\citet{ros12}.}
 \label{f:compare_f275w_f336w}
\end{figure*}

\begin{figure*}[!thb]
  \centerline{
       \epsfig{figure=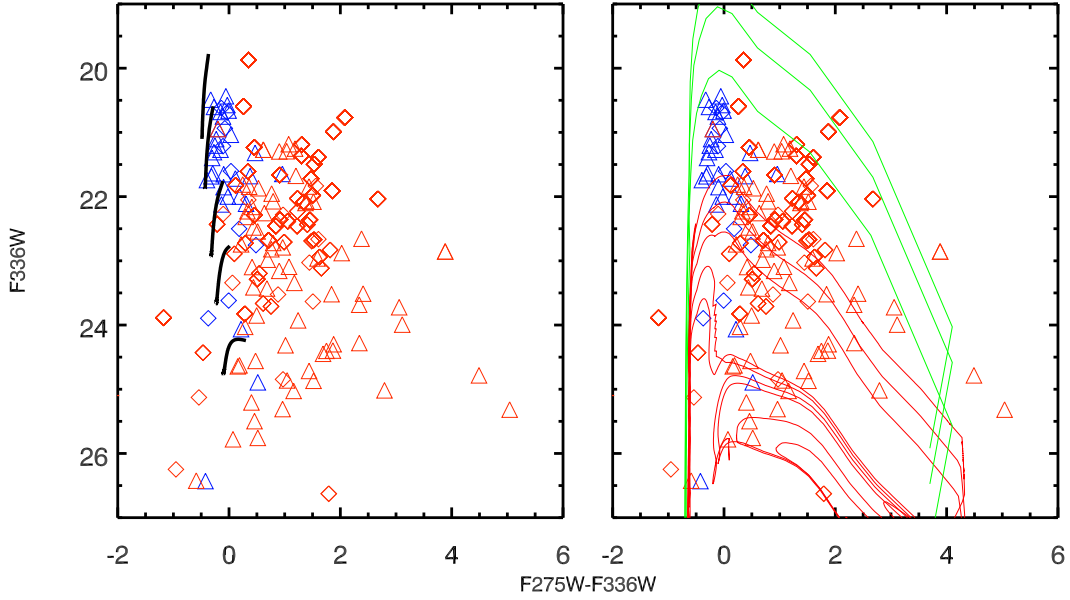,width=1.0\textwidth,angle=0}
       }
 \caption{The CMD (F275W-F336W vs F336W) of stars detected in the 
 HRC observation with F336W$<$23.3 mag. The blue and red symbols represent 
 the stars with F336W-F435W$<$-0.3 and F336W-F435W$>$-0.3 respectively, 
 while the diamonds and triangles inside and outside the galactocentric radius of 3\arcsec .
 In the left and right columns, we compared the CMD 
 with the MS tracks for young stars with 9, 7, 5, 4 and 3 solar mass (left) and 
 the tracks of the evolved low-mass stars at the P-AGB (green lines, from 
 top to bottom, 0.9, 0.75 and 0.6 $M_{\odot}$) and 
 HP-HB (red lines, from 
 top to bottom, 0.58, 0.56, 0.54, 0.5 and 0.48 $M_{\odot}$.) phases (right), as 
 described in~\citet{ros12}.}
 \label{f:hrc_compare_f275w_f336w}
\end{figure*}

First, like~\citet{ros12}, 
we analyzed the UV CMD  
for the WFC stars in the BP and the 
stars with F435W$<$23.3 mag in the HRC observation.
The \hst\ WFC3/UVIS F275W (2750\AA ) and F336W (3375\AA ) 
catalog of the PHAT survey
is from~\citet{wil14}. We used a 0.1\arcsec\ radius to identify the counterparts 
between the catalogs of~\citet{wil14} and our ones. %Because the FoV of the WFC3/UVIS 
%observation is smaller than that of the ACS/WFC~\citep{dal12}, 
%at $R$$>$230\arcsec , only 23\% of 
%stars in the BP in the Fig.~\ref{f:hess} have 
%counterparts in the UVIS observation. 
Fig.~\ref{f:compare_f275w_f336w} 
shows the UV CMD for the stars 
in the BP of Fig.~\ref{f:hess} at three different $R$ ranges  
defined in Fig.~\ref{f:young_dis} and Fig.~\ref{f:hrc_compare_f275w_f336w} 
gives the one in the HRC observation. % at 
%two ranges of galactocentric radii (inside or outside 3\arcsec ) and 
%different colours (F330W-F435W$<$-0.3 and F330W-F435W$>$-0.3). 
For the WFC observation, we know that the majority of the stars with 
$R$ $>$ 230\arcsec , are the young massive ones within the 5 kpc star formation ring. 
Therefore, it is not surprising that in the CMD, these stars follow a diagonal, 
which can be well explained by MS stars with mass from 4 to 9 $M_{\odot}$. 
Therefore, their age should be younger than 200 Myr.  
Instead, for the stars within $<$230\arcsec, they have a wide F275W-F336W color 
range. Because the extinction in the M31 bulge is very small 
($A_V$$<$0.15 mag,~\citealt{don16}, which corresponds to 
$A_{F275W}$-$A_{F336W}$$<$0.13 mag,~\citealt{don14}), 
the red stars can not be explained by MS stars 
with high extinction. Instead, these stars fall into the tracks of stars with 0.54 to 0.6 M$_{\odot}$. 
Therefore, these stars should be evolved low-mass stars. % in the post-AGB phases. 
In the HRC observation, 
the stars with F330W-F435W$<$-0.3 (blue symbols), also fall into 
a narrow F275W-F336W color range, like the BP stars in the WFC observation 
with $R$ $>$ 230\arcsec . On the other hand, the 
stars with F330W-F435W$>$-0.3 have a wide color range (red symbols), 
like the BP stars in the WFC observation 
with $R$ $<$ 230\arcsec\ and should be evolved low mass stars in the 
P-AGB, PE-AGB or AGB-manque phase. 

Second, in order to further exclude the possibility that the blue stars 
(F330W-F435W$<$-0.3) in the HRC observation are PNe, we cross-correlate 
our catalog with previous PNe catalogs.  
We used 1) the 12 PNe candidates in the central 21\arcsec\ 
($\sim$80 pc) reported by~\citet{pas13} using 
the SAURON integral-field spectroscopic 
observation; 2) the PNe candidates from Li Anqi, et al. (2018, in preparation). 
They analyzed the HST/UVIS observation from HST program, 
GO-12174~\citep{don14}, which includes one HST/UVIS pointing 
(2\arcmin$\times$2\arcmin , i.e.$\sim$450 pc$\times$450 pc)
centered 
on M31* with a narrow-band filter F502N (5013\AA ) 
and a medium-band filter F547M (5475\AA). 249 PNe candidates 
have been identified according to their unusually blue F502N-F547M 
colors, potentially due to the strong [\ion{O}{3}] $\lambda$5007\AA\ emission 
line, which is widely identified in PNe. 
%3) Beyond the FoV of HST/UVIS observation, we used 
%the PNe catalog from~\citet{mer06}, which used the Planetary Nebular 
%Spectrograph to identify 2615 PNe candidates in M31. 
We found that none of the blue stars in the HRC observation 
have counterparts in these two catalogs at a 
0.5\arcsec\ distance. Therefore, our results suggest that 
the blue stars in the HRC observation should not be 
PNe candidates, but instead should be young MS stars. 

%they can not be PNe candidates, 
%and instead should be young MS stars. 

%In summary, our results suggest that the blue stars in the HRC observation 
%are not evolved stars and could be young MS stars. 
%the M31 bulge 
%contain a non negligible amount of young MS stars. (the stars with blue F330W-F435W color). 

\subsection{The stellar populations of the M31 bulge}\label{ss:population}
In this subsection, we summarize the stellar populations in the M31 bulge. 

%with different ages, 1) $>$ 5 Gyr, 2) around 1 Gyr and 3) $<$ 500 Myr 
%in the M31 bulge. 
\subsubsection{Stars older than 5 Gyr}
As previously known~\citep{ols06,sag10}, the M31 bulge is dominated by 
stars older than 
$>$ 5 Gyr, which contribute more than 70\% of 
the total mass in the innermost annulus 
(see Fig.~\ref{f:cum_evo}. This is the lower limit, 
because of the detection limit). The old stars have super solar metallicity, 
with [Fe/H]$\sim$0.3 at all nine annuli. 

\subsubsection{Stars younger than 500 Myr}
Evidence of the existence of a $<$500 Myr old young 
stellar population has been shown in quantitative and qualitative studies  
above. First, the stars with 
F330W-F435W$<$-0.3 and F435W$<$23.3 mag have a very small 
F275W-F336W color range and can be explained by the 
tracks of young massive stars. Second, in \S\ref{ss:cmd}, we mentioned  
that the CMD evolves along $R$. In the 
inner region, there are more stars 
at 21$<$F814W$<$22 and 1$<$F475W-F814W$<$3, compared to 
the \#7 annulus.  
These stars are not due to 
the blending of dim stars in the M31 bulge, 
as demonstrated in \S\ref{ss:blending}. Their locations 
in the CMD are consistent 
with those predicted by a stellar population with around solar 
metallicity and $\sim$200 Myr 
old. Furthermore, the SFH derived from the 
CMD fitting (Fig.~\ref{f:sfh_evo} and Fig.~\ref{f:hrc_cmd_output}) 
also shows that there is 
a star formation activity less than 500 Myr ago with roughly solar 
metallicity, especially the HRC observation 
with fine sampling and less affect from 
the blending. Therefore, these stars could be the relatives of 
young blue stars identified in the 
central 0.25\arcsec\ by~\citet{lau12}. 
% and those in the HRC observations 
%(the central 14\arcsec ), 
%in the large scale. This result confirms previous works by 
%\citet{don16} and \citet{con16}. 

\begin{figure*}[!thb]
  \centerline{
       \epsfig{figure=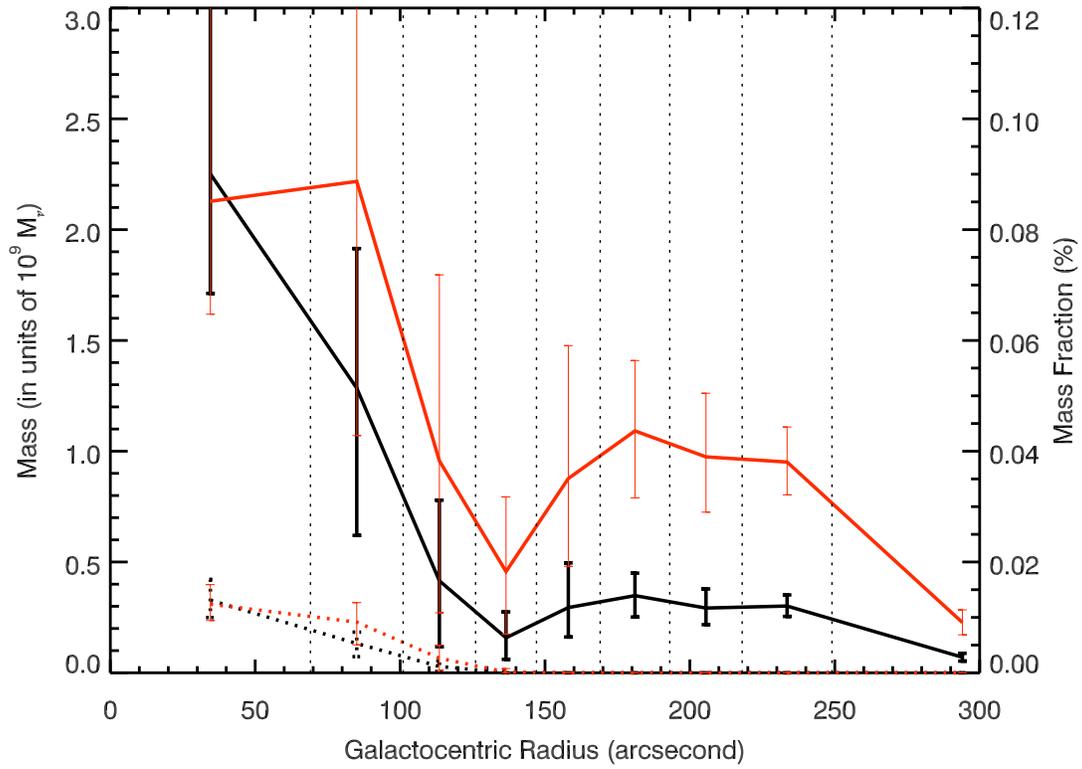,width=1.0\textwidth,angle=0}
       }
 \caption{The total mass (black lines) and mass fraction (red lines) of the stellar 
 population of $<$500 Myr old (dotted lines) and 900 Myr$<$Age$<$ 2 Gyr old (solid lines). 
 The eight 
 black vertical dotted lines are the boundaries of the nine annuli 
 defined in Table~\ref{t:annu}.}
 \label{f:young_inter_mass_evo}
\end{figure*}

%However, the mass of this stellar population is small, 
%4.9$^{+0.18}_{-0.16}$$\times10^8$ M$_{\odot}$, 
However, this stellar population only contribute 0.5\% of the 
total stellar mass in our FoV. %The latter is derived from 
%the \spitzer/IRAC 3.6 $\mu$m image (see \S\ref{ss:irac}). 
Fig.\ref{f:young_inter_mass_evo} gives the mass and the 
mass fraction of this young stellar population. 
The mass fraction of young stellar 
population is 1.2\%, then decreases to 0\% in the \# 5 annuli and beyond. 
%Considering the blending effect in the central annuli, the mass fraction 
%of the young stellar population should be even less. 
%In the HRC observation with high angular resolution, the star formation 

\subsubsection{Intermediate-age Stellar population ($\sim$1 Gyr)}
Fig.~\ref{f:sfh_evo} shows that in all the nine annuli, there is a SFR peak 
around 1 Gyr ago. The total mass of this stellar 
population (0.9 to 2 Gyr) is 5.4$^{+2.8}_{-2.0}$$\times10^9$ M$_{\odot}$, 5.4\% of the 
total stellar mass in our FoV. This stellar population contributes 9\% of 
the total stellar mass in the innermost annuli, but decreases to 
$\sim$4\% at \# 3 annuli. %We should be aware that in the 
%central two annuli, the blending could play an important role in the 
%apparent enhancement of star formation rate for 
%the activity $\sim$1 Gyr ago. 
The age of this stellar population 
evolves along $R$ too; %for example, %except for the central 
%two annuli, in which the blending effect is strong and there is contribution 
%from stellar population $<$ 500 Myr old, 
the mass-weighted age 
increases from 1 Gyr to 1.5 Gyr. On the other hand, the mass-weighted 
metallicity decreases along $R$. The $\sim$ 1 Gyr old stars can not have been 
contributed by the M31 disk, because no population with a similar age has 
been reported in previous deep \hst/ACS observations 
in the M31 disk~\citep{bro06}. 

\subsection{The SFH in the M31 bulge}\label{ss:formation}
Our studies above show that the star formation history in the M31 bulge 
is more complicated than we knew before. First, because the existence of 
a large number 
of old and metal-rich ([Fe/H]$>$0) stars, the 
majority of the M31 bulge could have formed in the major merger 
of several metal rich systems. 
On the other hand, recently,~\citet{bla17} 
suggested that 2/3 of the M31 bulge could come from 
a box/peanut bulge due to the secular process, 
from their N-body simulations.
%our new result is consistent 
%with previous studies~\citep{ols06,sag10} that the M31 bulge includes a lot of 
%metal-rich  old stellar population. Therefore, Then the star formation decreased  
%in the M31 bulge and the whole system stayed quiet till $\sim$2 Gyr ago. 
Second, the star formation rate increased quickly around 1 Gyr ago 
in all the annuli, which suggest that this star formation activity 
permeated the entire M31 bulge. 
Star formation activities with similar ages have not been seen in 
the M31 disk~\citep{bro06}. Considering the age and metallicity 
evolution along $R$, this population could
have arisen from secular processes; 
 % due to the secular 
%evolutionary process; 
when the molecular clouds spiraled into 
the nuclear region, the star formation activities first happened in the 
outer region of the M31 bulge. The supernova explosions 
of massive stars contaminated the remaining molecular clouds, 
which kept falling into the nuclear region to trigger more star 
formation. % indicates this star formation activities penetrate in 
%the M31 bulge. 
Third, 
the potential star formation activity less than 500 Myr ago identified in the 
M31 nucleus is 
only found in the central three annuli ($<$130\arcsec , $\sim$ 500 pc, see 
also~\citealt{lau12}), 
which means 
that this star formation event was rather localized.
%that, %unlike the 1 Gyr old star formation activity, 
%this star formation activity looks more 
%like a local one. 
Instead, if the interaction between M31 and the 
satellite galaxy, M32, as suggested by~\citet{blo06}, was real and triggered 
this star formation, we expect that the 200 Myr old stellar population  
should be distributed across a larger region
%should distribute 
%in a large field-of-view, 
which does not agree with the observation. 
Another 
possibility is that M31 used to  
have a central molecular zone like the one in the Milky Way. 
An intense star formation happened  
less than 500 Myr ago. Then the star formation activity %run out 
%of 
depleted the 
molecular clouds 
%gas
%the remaining gas 
and the feedback from the massive stars blew away
the remaining gas  
%the molecular clouds 
and stopped any further star formation activity. 

\subsection{Comparison with the SED fitting}\label{ss:com_sed}
The M31 bulge provides us a good lab to examine the reliability to 
use the SED fitting of integrated stellar light 
to get the SFH of remote galaxies, using the output of the CMD fitting 
as a candle. For example,~\citet{gar12} 
and~\citet{rui15} test the SFH derived by both methods and they 
conclude that the SED fitting is a reasonable tool to derive the 
general properties of unresolved stellar populations. 

\citet{don15} used the Starburst99 to perform the SED fitting of 
the photometric observation from mid-UV to NIR observation of the M31 
bulge. Limited by ten data points, they used only two stellar populations: one old 
metal-rich and another intermediate-age one. The age of
 the intermediate-age stellar population is 300 Myr in the central 5\arcsec . Then, it 
 increases to 600 Myr beyond that and slowly increases to 800 Myr at 
 $R$$\sim$180\arcsec . The metallicity of this population is 2.5Z$_{\odot}$ at the 
 central 5\arcsec and then decrease to $Z_{\odot}$. The mass fraction of this 
 stellar population is 0.2-2 percent. 
 
 %Our CMD fitting result is consistent with those of \citet{don15}. 
 Compared to~\citet{don15}, the advantage of the current study is that we 
 can get finer resolution in the age distribution. For example, below 2 Gyr, 
 we find that there are two significant star formation activities: one at around 
 1 Gyr and the other one at less than 500 Myr ago. The luminosity-weighted age 
 of the stellar populations $<$ 2 Gyr in our work is given as red lines 
 in Fig.~\ref{f:age_me_r_evo}. In the innermost annuli (\# 1), 
 our luminosity-weighted age is 900 Myr, which is larger than 
 300 Myr in the central 5\arcsec , but close to the age value beyond that 
 given in~\citet{don15}. The difference in the central 5\arcsec\ is because 
 1) limit by the number statistic and the detection limit 
 of the WFC observation, the CMD fitting method cannot divide the star 
 catalogs into small groups with the spatial resolution as high as that of the 
 SED fitting based on the 
 integrated light (5\arcsec ); 2) the mass fraction of the $<$500 Myr old 
 stellar population is low, which 
 cannot reduce the luminosity-weighted age. However, as mentioned 
 in \S\ref{ss:blueplume}, in the HRC observation, 
 we indeed find the candidates of massive stars with F330W-F435W$<$-0.3. 
 The total mass of this stellar population (4$\times10^5$ $M_{\odot}$ at the 
 375 Myr age bin) in the central 3\arcsec\ derived from the CMD fitting of the HRC 
 observation is close to 1$\times10^5$ $M_{\odot}$ (in the 0-5\arcsec ) given 
 in~\citet{don15}. The luminosity-weighted metallicity is 
 higher than that of the SED fitting, however, which is still consistent within the 
 large uncertainty, because in the SED fitting, we do not have any narrow-band filters 
 which are sensitive to the metallicity variation. Therefore, in one word, the SED fitting 
 with the integrated light can provide similar results of the CMD fitting, but 
 with higher angular resolution and poor age and metallicity resolution. Future SED fitting 
 of optical spectra derived from the CAHA PPAK observation of the M31 bulge 
 can provide us with more finer age and metallicity resolution to be compared with 
 the CMD fitting (Garc\'ia-Benito et al. in preparation).

\section{Summary}\label{s:summarize}
We have investigated the stellar populations in the central 5.5\arcmin\ of the M31 
bulge by using \hst/ACS WFC and HRC observations.  
Our field-of-view covers the whole M31 bulge and includes a small 
portion of the 5 kpc star formation ring in the north-west corner of our observation. 
We have produced Nyquist-sampled images and constructed %the 
%deconvolution 
source catalogs. The whole source catalog has been divided into nine 
annuli, centered on the M31*. We have performed the CMD fitting 
to study %the evolution of 
the stellar populations in 
the M31 bulge. 

The main results have been summarized below. 

\begin{itemize}
\item We have demonstrated that our deconvolution method on the 
Nyquist-sampled image performed better in the crowded field than 
the pipeline described in~\citet{dal12} with the`DOLPHOT' software; in 
the artificial star cluster test, the output catalogs derived by both methods 
are brighter and reddened than the input stellar populations; However, the 
CMD produced by the deconvolution method is closer to the input one. 

\item More than 2000 stars fall in the blue plume 
region in the WFC observation. Among them, the stars in the 
5 kpc star formation ring (galactocentric radius $>$ 230\arcsec ,$\sim$900 pc)
fall onto a diagonal line in the \hst/WFC3 UVIS 
F275W-F336W VS F336W CMD, %which is definitely  
which indicates that they are definitely 
main-sequence stars. However, 
the stars inside the central 230\arcsec\ are widely distributed in the 
F275W-F336W VS F336W CMD and could be explained 
by evolved low-mass stars, such as PAGB, PE-AGB or AGB-manque. 

\item In the HRC observation of the central 14\arcsec , the stars 
with F435W$<$23.3 mag could be divided 
into two parts by using F330W-F435W=-0.3. The F275W-F336W color of the 
blue one can be explained by main-sequence stars, while the red 
one could also be evolved low-mass stars. The former one confirms previous 
evidence that there is indeed an $\sim$ 200 Myr old 
intermediate-age stellar population in the M31 nucleus. 

\item From the CMD fitting of the WFC and HRC observations, we 
can obtain the following information about the star formation history 
in the M31 bulge: 1) Overall, the majority of the stars in the M31 
bulge, at least $>$ 70\%, are old ($>$ 5 Gyr) and metal-rich ([Fe/H]$>$0) stars. 
Subsequently,  
the star formation rate decreased till $\sim$ 2 Gyr. 2) %There is a burst of 
%star formation which happened $\sim$ 1Gyr in the whole M31 bulge. 
A star formation burst occurred about 1 Gyr ago throughout the entire bulge. 
3) The 
star formation activity less than 500 Myr ago with solar metallicity 
was only concentrated in 
the central $\sim$130\arcsec\ (500 pc). 

\item The identification of the 1 Gyr old star formation activity confirms 
our previous results derived from the spectral energy distribution fitting of the 
photometric observation at ten bands of the HST WFC3/ACS observation. 
%This stellar population widely distributed in the M31 bulge. 
%Their radial gradients 
%of age and metallicity suggests that they could be contributed the 
The radial gradients of age and metallicity of these stars suggest 
that they may have formed during the infall of molecular 
clouds from the M31 disk~\citep{opi18}. 

\item The spatial distribution of the 
stellar population less than 500 Myr old is %only limited in 
%the central $<$130\arcsec (i.e. $\sim$ 500 pc), but 
%not beyond that. This 
inconsistent with the scenario that the potential collision between 
the M31 and M32 triggered the star formation, which should produce star 
formation in a larger scale. Instead, we suggest that they could be due to 
the nuclear star formation, like the central molecular zone in the Milky Way. 
Then the molecular cloud has been %run out of due to the 
consumed by star formation activity 
and also been destroyed by the feedback. 

\item The properties of the stellar populations in the M31 bulge derived 
by the spectral energy distribution fitting of the integrated light is similar to 
those from the CMD fitting, which suggests that spectral energy distribution 
fitting method is still a reasonable method to study the stellar populations 
in remote galaxies.

\end{itemize}

\section*{Acknowledgments}
The research leading to these results has received funding from the
European Research Council under the European Union's Seventh Framework
Programme (FP7/2007-2013) / ERC grant agreement n$^{\odot}$ [614922]. 
and it was also supported by NASA
via the grant GO-12055, provided by the Space Telescope Science
Institute. This work
uses observations made with the NASA/ESA Hubble Space Telescope and
the data archive at the Space Telescope Science Institute, which is
operated by the Association of Universities for Research in Astronomy,
Inc. under NASA contract NAS 5-26555.  
H. D. acknowledges the hospitality of University of Massachusetts, Amherst 
during his visit. We are grateful to Q. Daniel, Wang and Philip Rosenfield for
many valuable comments and discussion and Pauline Barmby 
for providing the Spitzer/IRAC image.

%\begin{figure*}[!thb]
%  \centerline{
%       \epsfig{figure=fig/hrc_wfc_comp.ps,width=1.0\textwidth,angle=0}
%       }
% \caption{The comparison of the HRC F435W (left) and WFC F475W (right) observations  
% of a region in the celestial north of the M31 nucleus.}
% \label{f:hrc_wfc_comp}
%\end{figure*}

%\begin{figure*}[!thb]
%  \centerline{
%       \epsfig{figure=fig/fake_source.ps,width=1.0\textwidth,angle=0}
%       }
% \caption{\hst\ \acs\ F814W image of Brick 17 Field 16, with the artificial star
%   clusters marked with circles and corresponding IDs. The artificial star
%   clusters have different total mass and compactness.}
% \label{f:fake_source}
%\end{figure*}

%\begin{figure*}[!thb]
 % \centerline{
  %     \epsfig{figure=fig/hess_pub.ps,width=1.0\textwidth,angle=0}
  %     }
 %\caption{The Hess representation of the CMD at nine annuli in the M31
% Bulge. }
 %\label{f:hess_pub}
%\end{figure*}

\clearpage

\begin{deluxetable}{cccc}
  \tabletypesize{\normalsize}
 \tablecolumns{4}
  \tablecaption{Broad-band Filters}
  \tablewidth{0pt}
  \tablehead{
  \colhead{Filter} & 
  \colhead{Central Wavelength (\AA )} &
  \colhead{Width (\AA )} & 
  \colhead{Exposure$^a$} \\
 }
  \startdata
F475W & 4760 & 1458 & 10+\textbf{360}+\textbf{360}+\textbf{470}+\textbf{700} \\
F814W & 8333 & 2511 & 15+350+\textbf{800}+\textbf{550} \\
F330W & 3354 & 588  & 8$\times$298 \\
F435W & 4297 & 1038 & 672$\times$4+679$\times$8\\ 
\enddata
\tablecomments{$^a$: the solid numbers mean that the corresponding
  dithered exposures are used to produce the Nyquist Images for the
  WFC observations in \S\ref{s:reduction}.}
\label{t:filters}
\end{deluxetable}

\begin{deluxetable}{cccccc}
\tabletypesize{\small}
 \tablecolumns{6}
  \tablecaption{Completeness limit}
\tablewidth{0pt}
  \tablehead{
  \colhead{ID} & 
  \colhead{Radius} &
  \colhead{F475W(70\%)} &
  \colhead{F475W(50\%)} &
  \colhead{F814W(70\%)} &
  \colhead{F814W(50\%)} \\
}
\startdata
1&0\arcsec -69\arcsec &24.2&24.6&21.9&22.3\\
2&69\arcsec -101\arcsec &24.6&25.0&22.4&22.8\\
3&101\arcsec -126\arcsec &24.8&25.2&22.7&23.1\\
4&126\arcsec -147\arcsec &25.0&25.4&22.9&23.3\\
5&147\arcsec -169\arcsec &25.2&25.5&23.1&23.5\\
6&169\arcsec -193\arcsec &25.3&25.7&23.2&23.7\\
7&193\arcsec -218\arcsec &25.4&25.8&23.4&23.9\\
8&218\arcsec -249\arcsec &25.6&26.0&23.6&24.0\\
9&249\arcsec -339\arcsec &25.9&26.3&23.9&24.4\\
\enddata
\tablecomments{The 70\% and 50\% completeness limit of the 
WFC observation at nine annuli of the F475W and F814W bands.}
\label{t:annu}
\end{deluxetable}

\begin{deluxetable}{ccccccc}
  \tabletypesize{\footnotesize}
 \tablecolumns{7}
  \tablecaption{Artificial Star Clusters}
  \tablewidth{0pt}
  \tablehead{
  \colhead{} & 
  \colhead{log Age} &
  \colhead{Total Mass} & 
  \colhead{Radius$^a$} &
\colhead{MD$^b$} &
\colhead{SB$^c$} & 
 \colhead{$A_V$}\\
  \colhead{ID} & 
  \colhead{(yr)} & 
  \colhead{(M$_{\odot}$)} & 
  \colhead{(pc)} & 
  \colhead{ (M$_{\odot}$  pc$^{-2}$)} & 
  \colhead{ (mag arcsec$^{-2}$)} &
  \colhead{(mag)} \\
 }
  \startdata
1& 9.6&   9.6e+04&3.2&3055.8&18.0&   0.2\\
2& 7.8&   2.7e+03&3.0&  95.0&19.6&   0.3\\
3& 7.8&   3.5e+03&2.7& 150.6&19.0&   0.6\\
4& 9.4&   7.5e+03&1.5&1126.8&20.1&   1.5\\
5& 8.1&   6.2e+02&3.2&  19.3&21.0&   0.5\\
6& 9.8&   9.7e+03&3.1& 313.4&19.6&   0.3\\
7& 6.7&   8.6e+02&2.7&  38.4&20.0&   0.4\\
8& 8.4&   1.8e+03&8.8&   7.3&20.6&   1.6\\
9& 8.1&   5.4e+02&1.9&  45.0&20.7&   0.3\\
10& 7.3&   5.8e+02&2.7&  25.5&20.6&   1.8\\
11& 9.6&   1.8e+04&5.8& 171.7&19.4&   0.9\\
12& 8.6&   1.2e+03&2.3&  74.3&20.4&   0.2\\
13& 9.7&   9.3e+03&4.7& 133.7&21.0&   0.8\\
14& 9.9&   5.0e+03&2.7& 219.0&21.0&   1.3\\
15& 9.6&   2.0e+04&1.3&3926.0&19.7&   2.1\\
16& 9.6&   4.0e+04&3.7& 927.5&19.1&   0.3\\
17& 9.9&   6.5e+04&2.8&2699.5&20.8&   2.9\\
18& 8.1&   3.5e+02&3.1&  11.4&20.5&   1.6\\
19& 9.7&   9.1e+04&1.8&8511.0&20.9&   2.8\\
20& 7.1&   1.1e+03&7.0&   7.4&20.0&   0.9\\
21& 7.8&   3.5e+03&1.6& 442.6&20.9&   1.4\\
22& 7.4&   2.5e+02&1.4&  40.2&19.7&   1.3\\
23& 8.2&   1.8e+03&3.6&  43.4&20.6&   0.8\\
24& 8.6&   1.4e+04&1.9&1198.3&18.2&   1.1\\
25& 8.6&   7.7e+02&1.7&  87.3&20.6&   1.3\\
26& 6.8&   3.3e+02&1.8&  31.6&20.0&   1.4\\
27& 8.8&   2.4e+03&1.6& 278.4&20.2&   1.4\\
28& 8.6&   7.8e+03&2.8& 307.2&20.9&   0.3\\
\enddata
\tablecomments{$^a$: the effective radius. $^b$: the mass surface density within the effective radius. $^c$: 
the F814W surface brightness within the central 0.5\arcsec\ (1.9 pc).}
\label{t:fake}
\end{deluxetable}

\end{document}